\documentclass[twoside,11pt]{article}

\usepackage{subcaption}

\usepackage[preprint]{jmlr2e}

\usepackage[linesnumbered]{algorithm2e}
\usepackage{amsmath}
\usepackage[noabbrev,capitalise]{cleveref}
\usepackage{colortbl}
\usepackage{multirow}
\usepackage{physics}

\def\bx{\textbf{x}}

\def\sH{\mathsf H}

\jmlrheading{0}{0000}{00-00}{00/00}{00/00}{paperId}{Enrico Zardini, Enrico Blanzieri and Davide Pastorello}

\ShortHeadings{Implementation and Evaluation of a QML Pipeline}{Zardini, Blanzieri and Pastorello}
\firstpageno{1}

\usepackage{xy}
\xyoption{matrix}
\xyoption{frame}
\xyoption{arrow}
\xyoption{arc}

\usepackage{ifpdf}
\ifpdf
\else
\PackageWarningNoLine{Qcircuit}{Qcircuit is loading in Postscript mode.  The Xy-pic options ps and dvips will be loaded.  If you wish to use other Postscript drivers for Xy-pic, you must modify the code in Qcircuit.tex}
\xyoption{ps}
\xyoption{dvips}
\fi

\entrymodifiers={!C\entrybox}

\newcommand{\qw}[1][-1]{\ar @{-} [0,#1]}
\newcommand{\qwx}[1][-1]{\ar @{-} [#1,0]}

\newcommand{\gate}[1]{*+<0.5em>{#1} \POS ="i","i"+UR;"i"+UL **\dir{-};"i"+DL **\dir{-};"i"+DR **\dir{-};"i"+UR **\dir{-},"i" \qw}

\newcommand{\control}{*!<0em,.095em>-=-<.2em>{\bullet}}

\newcommand{\ctrl}[1]{\control \qwx[#1] \qw}

\newcommand{\targ}{*+<.04em,.04em>{\xy ="i","i"-<.50em,0em>;"i"+<.50em,0em> **\dir{-}, "i"-<0em,.47em>;"i"+<0em,.47em> **\dir{-},"i"*\xycircle<.6em>{} \endxy} \qw}
\newcommand{\qswap}{*=<0em>{\times} \qw}
\newcommand{\lstick}[1]{*!R!<.5em,0em>=<0em>{#1}}

\newcommand{\Qcircuit}{\xymatrix @*=<0em>}

\begin{document}

\title{Implementation and Empirical Evaluation of a \\ Quantum Machine Learning Pipeline \\
for Local Classification}

\author{\name Enrico Zardini$^{1}$ \email enrico.zardini@unitn.it \\ 
        \name Enrico Blanzieri$^{1,2}$ \email enrico.blanzieri@unitn.it \\ 
        \name Davide Pastorello$^{1,2}$ \email d.pastorello@unitn.it \\ 
        \addr $^1$ Department of Information Engineering and Computer Science, University of Trento, via Sommarive 9, 38123 Povo, Trento, Italy \\
        \addr $^2$ Trento Institute for Fundamental Physics and Applications, via Sommarive 14, 38123 Povo, Trento, Italy}

\editor{Editor Names (Name and Surname)}

\maketitle

\begin{abstract}
In the current era, quantum resources are extremely limited, and this makes difficult the usage of quantum machine learning (QML) models. Concerning the supervised tasks, a viable approach is the introduction of a quantum locality technique, which allows the models to focus only on the neighborhood of the considered element. A well-known locality technique is the $k$-nearest neighbors ($k$-NN) algorithm, of which several quantum variants have been proposed. Nevertheless, they have not been employed yet as a preliminary step of other QML models, whereas the classical counterpart has already proven successful. In this paper, we present (i) an implementation in Python of a QML pipeline for local classification, and (ii) its extensive empirical evaluation. Specifically, the quantum pipeline, developed using Qiskit, consists of a quantum $k$-NN and a quantum binary classifier. The results have shown the quantum pipeline’s equivalence (in terms of accuracy) to its classical counterpart in the ideal case, the validity of locality’s application to the QML realm, but also the strong sensitivity of the chosen quantum $k$-NN to probability fluctuations and the better performance of classical baseline methods like the random forest.
\end{abstract}


\section{Introduction}
\label{sec:introduction}
The application of locality to the quantum realm is extremely relevant. Indeed, currently, i.e., in the noisy intermediate-scale quantum (NISQ) era \citep{preskill2018quantum_nisq}, the available quantum devices are limited in the number of qubits and the fidelity of gates. The qubit is the quantum analogue of a classical bit; in 2019, Google claimed to have reached quantum supremacy, i.e., an advantage with respect to classical computers, using a processor with 54 qubits \citep{google_qsupremacy}, a very large number compared to the chips accessible for practical purposes nowadays. Nevertheless, the performance in the task considered by Google was still very low. Regarding the fidelity of gates, it is always undermined by the presence of intrinsic fabrication issues since these universal quantum architectures are based mainly on superconducting circuits. Specifically, the gate fidelity represents how much the operation performed by a gate is close to the ideal one \citep{Magesan_2011_gate_fidelity}. This measure also worsens by increasing the number of qubits due to the lack of all-to-all connections between them. 

In this scenario, the usage of quantum machine learning (QML) models turns out to be difficult: it is not possible to encode large quantities of data in the quantum device due to the low number of qubits (typically, also a data index must be encoded); in addition, the presence of noise in the execution of quantum operations limits the quality of results, especially if a consistent amount of samples is considered. Several quantum machine learning models have already been proposed, either classical-quantum hybrid \citep{Havlicek2019_hybrid_svms} or entirely quantum \citep{quantum_svm}. The latter type is, of course, the most interesting one since the models can fully exploit the quantum potentialities without the need to interact with classical procedures; at the same time, these models suffer more from the aforementioned issues. For instance, the quantum support-vector machine (SVM) proposed by \citet{quantum_svm} has been implemented and tested by Z.~\citet{quantum_svm_implementation}, but the task considered was very small.

In conclusion, the possibility of reducing the number of qubits required to solve a problem is extremely relevant in the NISQ era. In this way, it becomes feasible to address bigger, and thus more significant, problems and concretely exploit the potentialities of quantum machine learning. Focusing on the supervised tasks, the introduction of a quantum locality technique represents a valid direction in this sense. Indeed, by looking at only the neighbourhood of the considered element, it is possible to reduce the number of samples that the quantum models must process. An alternative approach could be reducing the number of qubits required to encode each sample by using an autoencoder or a dimensionality reduction technique such as the singular value decomposition (SVD). Actually, a classical-quantum hybrid version of these two techniques has already been proposed by \citet{Romero_2017_quantum_autoencoder} and X. \citet{Wang2021variationalquantum}, respectively.

While the reduction of the number of input samples to a quantum machine learning model through a quantum locality technique has not been addressed yet in the literature, the classical counterpart has already been investigated and has proven successful. For instance, a local SVM trained on the samples selected by a $k$-nearest neighbors ($k$-NN) model has been presented and tested by \citet{local_svm}, with good results. Actually, the simplest and most effective locality technique is represented precisely by the $k$-NN, which picks out the elements closest to the target one according to a given metric. Different quantum variants of the $k$-NN algorithm have been proposed; moreover, an interesting application of a quantum $k$-NN version in combination with the Grover search algorithm \citep{grover} has also been presented by \citet{quantum_reccomendation_sys}, namely, a quantum recommendation system. 

Eventually, the empirical evaluation has always been an essential part of the machine learning research paradigm, with the UCI Machine Learning Repository \citep{uci_ml_repo} playing a major role in setting de facto standard benchmarks for the experimental assessment of machine learning (ML) algorithms. Recently, massive benchmark datasets have been proposed and used for studying and advancing the state of the art of deep learning neural networks \citep{deep_learning_nn_datasets}. Instead, the area of quantum machine learning has not yet matured enough to produce established benchmarks; this is due to the mainly theoretical nature of the research in the current phase, as well as the limited dimension and reliability of the available machines. Nevertheless, a fair and systematic comparison between quantum and classical machine learning is necessary in order to progress. To this end, a platform that supports the execution of both kinds of algorithms, and provides access to the existing quantum devices and/or their simulators, is highly desirable.

In this work, we present the implementation and the empirical evaluation of a quantum pipeline consisting of a quantum $k$-nearest neighbors algorithm~\citep{qknn_implemented} and a quantum cosine-based binary classifier~\cite{binary_classifier_published}. The code, which is publicly accessible,\footnote{The code is available at \url{https://github.com/ZarHenry96/quantum-ml-pipeline}.} integrates access to available quantum computing resources, namely, IBM quantum computers, provides for testing on several UCI datasets whose dimension is compatible with the current quantum devices, and allows for comparisons with classical competitors. However, the quantum pipeline has not been tested on real quantum devices due to the retirement of the quantum computer of interest (the only one available with a free account that had enough qubits). In particular, the features supplied by the code permit to evaluate the role that locality can have in quantum machine learning.

The remainder of the paper is structured as follows: \cref{sec:background} presents some background information, \cref{sec:quantum-pipeline} describes the quantum pipeline, its implementation and complexity, \cref{sec:empirical-eval} deals with the experimental evaluation and the results obtained, and \cref{sec:conclusion} provides the conclusions.

\section{Background}
\label{sec:background}
This section provides background information about quantum computing, quantum machine learning, the classical and the quantum versions of the $k$-nearest neighbors algorithm, and the quantum binary classifier proposed by \citet{binary_classifier_published}.

\subsection{Quantum Machine Learning}
\label{subsec:quantum-machine-learning}
Quantum computing is a type of computation in which quantum phenomena, such as state superposition and entanglement, are exploited to perform calculations. It is the most prominent application of quantum information theory and delivers algorithms to solve efficiently problems that are hard for classical computers \citep{nielsen00}. 

Quantum machine learning is an emerging research area related to quantum computing. Basically, in QML, quantum computing techniques are applied to machine learning tasks in order to pursue computational advantages given the present context of ever-growing amounts of data to manage. In detail, QML algorithms may present relevant benefits in time and space complexity w.r.t classical ML algorithms. Remarkable examples are the embedding of quantum subroutines into ML schemes to efficiently calculate distances in the feature space \citep[e.g.,][]{Schuld_2017}, with advantages in classification and clustering, or Grover-based subroutines to find an item in an unsorted database \citep[e.g.,][]{q_nearest_neighbor}, with a quadratic speedup w.r.t. an exhaustive search. The latter are employed, for instance, in pattern recognition. The first proposals of quantum versions of ML algorithms were presented about twenty years ago \citep[e.g.,][]{trugenberger_q_sum,pattern_rec}, but the real interest in QML has sparked only in the last decade thanks to the development of the first available working prototypes of quantum machines like those manufactured by IBM \citep{IBM}, Rigetti \citep{Rigetti}, and D-Wave Systems \citep{D-Wave}, and the publication of many interesting results on quantum machine learning algorithms \citep[e.g.,][]{quantum_svm, Dunjko, biamonte, Havlicek2019_hybrid_svms}.

Within the quantum circuit model, one of the most popular, the basic notion of quantum computation is represented by the qubit, whose state is described by a unit vector $\ket\psi=\alpha\ket 0+\beta\ket 1$ in a two-dimensional complex Hilbert space, of which $\ket 0 $ and $\ket 1$ form an orthonormal basis. In particular, $\ket 0 $ and $\ket 1$ identify the vectors of the standard basis of $\mathbb C^2$. Instead, the absolute squares of the amplitudes $\alpha,\beta\in\mathbb C$ correspond to the probabilities of measuring the qubit in states 0 and 1, respectively; hence, $|\alpha|^2+|\beta|^2=1$. After a measurement process, the state of a qubit collapses to the post-measurement state, either $\ket 0$ or  $\ket 1$, according to the obtained outcome. In addition, the time evolution of isolated quantum systems (such as the qubits) is mathematically described by unitary operators, which are called quantum gates in the language of quantum computing. An example of a quantum gate acting on a single qubit is the Hadamard gate, whose action on the basis states is given by $H\ket 0=\ket +$ and $H\ket 1 =\ket -$, where $\ket\pm=1/\sqrt 2 \cdot (\ket 0\pm\ket 1)$. The corresponding matrix representation and the circuital symbol are
$$
H=\frac{1}{\sqrt 2}\left(\begin{array}{cc}
1 & 1\\
1 & -1
\end{array}\right)\qquad \text{and} \qquad
\Qcircuit @C=1.0em @R=2.0em {
&\gate{H}&\qw
}\quad. 
$$
Another important quantum gate is the controlled NOT gate (or CNOT), which operates over two qubits and acts as follows w.r.t. the computational basis:
$$
\Qcircuit @C=1.0em @R=2.0em {
\ket x& &\ctrl{1}&\qw& \ket x &\\
\ket y& &\targ&\qw & &\ket{x\oplus y} \,,
}
$$
where $x,y\in\{0,1\}$ and $\oplus$ is the sum modulo 2. In particular, by combining three CNOTs, it is possible to build the SWAP gate, a 2-qubit gate that swaps the qubits provided as input; its circuital definition is
$$
\Qcircuit @C=2em @R=2.8em {
& \qswap & \qw& \\
& \qswap\qwx & \qw &
} 
\Qcircuit @C=2em @R=1.5em {
&  & \\
& \lstick{:=}&\\
&  &
}
\Qcircuit @C=1em @R=1.5em {
& \targ&\ctrl{1} &\targ&\qw \\
& \ctrl{-1}&\targ &\ctrl{-1}&\qw &.
}
$$
Instead, the controlled version of the SWAP gate (a 3-qubit gate) is called Fredkin gate as its classical version, which is universal for classical reversible computation, and the corresponding circuital symbol is 
$$
\Qcircuit @C=2em @R=1.5em {
&\ctrl{1} &\qw&\\
& \qswap & \qw& \\
& \qswap\qwx & \qw &. 
}
$$
\vspace{0.3cm}

As an axiom of quantum mechanics, a system of $n$ qubits is described in the space $(\mathbb C^2)^{\otimes n}$. As a consequence, the dimension of the space where we can represent the data grows exponentially in the number of qubits; this is one of the main advantages of quantum computations. In practice, quantum algorithms are developed by composing the available quantum gates in order to produce a quantum state that encodes the solution of a given problem, and the readout is performed by measuring the quantum output state in the computational basis. Since the result of a quantum computation is probabilistic in general, a quantum algorithm must be repeated several times in order to provide a meaningful result. One of the main examples of the efficiency of quantum computation is the celebrated Shor’s algorithm \citep{Shor}, which solves the integer factoring problem (that is generally suspected to be not in \textbf{P}) in polynomial time. 

In QML, quantum algorithms are developed to solve machine learning tasks like classification, clustering, and pattern recognition \citep{schuld}. To this end, a relevant question is the representation of classical data into quantum states. The standard encoding used in quantum computing is the so-called basis encoding: binary strings $(x_1,...,x_n)$, with $x_i\in\{0,1\}$ and $i=1,...,n$, are translated into states of $n$ qubits $\ket{x_1,...,x_n}$ belonging to the basis of the $n$-qubit Hilbert space. Alternatively, many QML algorithms exploit the amplitude encoding, in which a classical data instance $\bx\in\mathbb R^d$ is encoded into the quantum superposition state $\ket \bx=\parallel\bx \parallel^{-1}\sum_{i=1}^d x_i \ket i$ of $\log_2 d$ qubits. Eventually, it is also worth presenting a simple example of quantum processing that is rather useful in QML (and is applied within the $k$-NN and the binary classifier considered in this work), i.e., the SWAP test \citep{swap_test}. The corresponding circuit is the following:
$$\label{swap}
\Qcircuit @C=1.0em @R=2.0em {
\ket 0\qquad&\gate{H}& \ctrl{1} & \gate{H}& \qw\\
\ket{\psi}\qquad&\qw& \qswap & \qw& \qw\\
\ket{\varphi}\qquad&\qw& \qswap\qwx & \qw &\qw \quad,
} 
$$ \\
where $\ket{\psi}$ and $\ket{\varphi}$ are $n$-qubit states. In detail, the SWAP gate, which acts on $\ket{\psi}$ and $\ket{\varphi}$, is controlled by the qubit initially prepared in $\ket 0$; this can be implemented through $n$ Fredkin gates. A simple calculation shows that the probability of measuring the value 0 in the first qubit is $\mathbb P(0)=1/2 \cdot (1+|\langle\psi|\varphi\rangle|^2)$. In addition, the estimation of $\mathbb P(0)$ up to an error $\epsilon$ requires $O(\epsilon^{-2})$ repetitions as given by the binomial proportion confidence interval for a Bernoulli trial. In practice, the SWAP test allows the efficient computation of the fidelity of the quantum states $\ket{\psi}$ and $\ket{\varphi}$, with the fidelity being defined for two pure quantum states as
\begin{equation}
    \label{eq:fidelity}
    \mathcal F(\ket\psi,\ket\varphi)= |\bra{\psi}\ket{\varphi}|^2 = (\cos(\psi, \varphi)\ \cdot \parallel \psi \parallel \cdot \parallel \varphi \parallel)^2 = \cos^2(\psi, \varphi)\ ,
\end{equation}
where $\cos(\psi, \varphi)$ is the cosine similarity of $\ket{\psi}$ and $\ket{\varphi}$, and the norms of $\ket{\psi}$ and $\ket{\varphi}$ are 1 by definition. Therefore, by encoding data vectors into the amplitudes of $\ket{\psi}$ and $\ket{\varphi}$, it is possible to compute their dot product and, thus, their cosine similarity (and distance) through the SWAP test.

\subsection{K-NN and Quantum K-NN(s)}
\label{subsec:k-nn-n-quantum-k-nns}
The $k$-nearest neighbors algorithm \citep{k_nn} is a really simple classification algorithm, and consists of three steps: \begin{itemize}
    \item the computation of the chosen distance metric between the test element and all training data points;
    \item the extraction of the $k$ elements closest to the test instance, namely, the $k$ nearest neighbors;
    \item the assignment of the class label through a majority voting based on the labels of the $k$ nearest neighbors (in the case of a classification task).
\end{itemize}
Instead, the quantum counterpart of the algorithm, of which several variants have been proposed, includes an additional step at the beginning, i.e., the preparation of data in a superposition state. This stage allows performing parallel operations, such as computing the distance of the test instance with respect to all training elements simultaneously (quantum parallelism).

Regarding the quantum $k$-NN variants, a conceptually simple one (but not so efficient) is described in the work of \citet{basic_qknn} and consists of two steps: the SWAP test algorithm is exploited to compute the distance (the Euclidean distance, in this case) between feature vectors, which are encoded in the amplitudes of a superposition state; a quantum minimization algorithm based on the Grover search \citep{durr_minimization}, also known as D\"{u}rr's algorithm, is used to find the $k$ nearest neighbors. In particular, each of the two steps requires multiple iterations with final measurements. 

A more complex variant has been presented by both \citet{qknn_img_class} and Y.~\citet{qknn_img_digit_rec}, and applied to the image classification task. The workflow is the following: the features are encoded as amplitudes of a quantum superposition state; the distance between test and training instances is computed through a SWAP test, but without measurements; the amplitude estimation (AE) algorithm \citep{ampl_ampl_and_estim} is then used to transfer the distance values encoded in amplitudes to qubit states \citep[measurements are not strictly necessary in this step, as shown by][]{q_nearest_neighbor}; finally, D\"{u}rr's algorithm is exploited to find the indices of the $k$ elements with minimum distance with respect to the test instance. It is worth highlighting that both AE and D\"{u}rr's algorithms are quite complex and include an oracle, i.e., they are based on a black-box function. Moreover, a very similar workflow is present in the work of \citet{q_nearest_neighbor}, although it is used for finding only the nearest neighbor.

Quantum $k$-NN versions employing a different metric, namely, the Hamming distance, have been proposed by \citet{ruan2017_qknn_hamming} and J.~\citet{qknn_hamming_binary}. Due to the metric chosen, the features must be expressed as bit strings; indeed, the Hamming distance represents the number of positions at which two strings differ. The advantage is a straightforward mapping to quantum states (basis encoding). In detail, the two considered $k$-NN variants share the initial steps: a superposition of the features quantum states is prepared; the difference between corresponding qubits, in training and test features, is computed through CNOT gates; the Hamming distance, which corresponds to the sum of the differences, is obtained by using the \textit{incrementation} circuit presented by \citet{incrementation_circuit}. Then, in the first variant \citep{ruan2017_qknn_hamming}, the training instances with a distance lower than a given threshold value are selected by means of an OR gate and a projection operation (actually, to do this, the qubits differences are reversed before the summation). Thus, there is no $k$ parameter and the number of nearest neighbors selected depends on the threshold value. Instead, in the second variant \citep{qknn_hamming_binary}, a novel quantum search procedure inspired by a binary search is applied to the distances in order to find the minimum value. By iterating it and removing each time the current minimum, the $k$ nearest neighbors are selected. The Hamming distance (computed with the procedure just described) is used also in the work of \citet{qknn_img_class_kl_transf} for image classification, but the search for the $k$ minimum distance values is performed through D\"{u}rr's algorithm.

Another method for computing the Hamming distance is exploited in the quantum $k$-NN variants presented by \citet{schuld_qknn_pattern_class} and \citet{qknn_herm_matrix_rec}. In detail, instead of summing up the qubits differences through the incrementation circuit, a unitary operation is applied to them in order to encode the values of the sums as quantum state amplitudes. This idea has been proposed first by \citet{trugenberger_q_sum}. After an ancillary qubit measurement, which is required to select the good amplitudes distribution (higher probabilities for lower distances), the classification is performed directly in both works without explicitly selecting the $k$ nearest neighbors. However, it is possible to identify the neighbors by repeating the entire process multiple times and measuring the post-ancillary-measurement state (instead of executing the classification step).

The last interesting quantum $k$-NN variant has been presented by both \citet{qknn_implemented} and \citet{qknn_cat_tensor_networks}. In particular, after the encoding of the data features as amplitudes of a superposition state, a SWAP test is performed. Then, the state of an ancillary qubit and of a qubit register (array), which indexes the training data, is measured. By iterating the procedure just described, it is possible to estimate a quantity proportional to the fidelity \citep{q_states_fidelity} between the training data and the test instance states and, therefore, find the $k$ nearest neighbors. Indeed, the fidelity corresponds to the squared scalar product for pure quantum states (see Equation \ref{eq:fidelity}). It is worth highlighting that, as shown by \citet{qknn_cat_tensor_networks}, multiple test instances can be processed in parallel by introducing an additional index register for the test data and putting the test instances in superposition (as the training ones). Actually, \citeauthor{qknn_implemented} have recently proposed also another quantum $k$-NN variant \citep{basheer2021quantum}: it exploits a generalization of D\"{u}rr's algorithm to find the indices of the $k$ nearest neighbors given a quite complex oracle as input. However, the resulting workflow is not so different from that of other previously described works. Basically, the oracle in question includes the SWAP test, a quantum analog-to-digital conversion algorithm \citep{q_analog_to_dig_conv} based on the phase estimation algorithm \citep{phase_estimation}, and some quantum arithmetic.

\subsubsection{A Quantum K-NN Variant in Detail}
\label{subsubsec:quantum-knn-basheer}
Let us describe more in detail the quantum $k$-NN algorithm proposed by \citet{qknn_implemented}. In order to do so, let us consider the dataset $\{\bx_i\}_{i={0,...,N-1}}$, with $\bx_i\in\mathbb R^d$, the test data instance $\bx\in\mathbb R^d$, and the fidelity, i.e., the squared cosine similarity (see Equation \ref{eq:fidelity}), as a distance measure. Within the amplitude encoding, the cosine similarity between $\bx_i$ and $\bx$ is nothing but the inner product $\langle\bx_i|\bx\rangle$ between the corresponding quantum states. In addition, let us assume that $N$ and $d$ are powers of 2 without loss of generality. Then, let us consider an index register of $\log_2 N$ qubits, where the indexes of the training instances are stored within the basis encoding, two $n$-qubit registers (with $n=\log_2 d$), where data are encoded into the amplitudes of the quantum states, and an ancillary qubit. The four registers are initialized in the state
\begin{equation}\label{knn1}
\frac{1}{\sqrt N}\sum_{i=0}^{N-1} \ket i\ket{\bx_i}\, \ket{\bx}\,\ket 0\in \sH_{index}\otimes\sH_n\otimes\sH_n\otimes\sH_a.
\end{equation}
In the state (\ref{knn1}), the superposition of the training data and the test instance are stored in two different registers. Now, let us perform the SWAP test on the two $n$-qubit registers controlled by the ancillary qubit, obtaining the state
\begin{equation*}
\ket{\Psi}=\frac{1}{2\sqrt N} \sum_{i=0}^{N-1} \ket i \left[(\ket{\bx_i}\ket{\bx}+\ket{\bx}\ket{\bx_i})\ket 0+(\ket{\bx_i}\ket{\bx}-\ket{\bx}\ket{\bx_i})\ket 1\right].
\end{equation*}
The probability of getting the outcome $\alpha\in\{0,1\}$ by a measurement process on the ancillary qubit is given by
\begin{equation*}
\mathbb P(\alpha)=\frac{1}{2}+(-1)^{\alpha}\frac{1}{2N}\sum_{i=0}^{N-1}|\langle\bx_i|\bx\rangle|^2,
\end{equation*}
and the corresponding post-measurement state stored in the four registers is
\begin{equation*}
\ket{\Psi_\alpha}=\frac{\sum_{i=0}^{N-1} \ket i(\ket{\bx_i}\ket{\bx}+(-1)^{\alpha}\ket{\bx}\ket {\bx_i})}{\sqrt {2\left(N+(-1)^\alpha\sum_{i=0}^{N-1} |\langle\bx_i|\bx\rangle|^2\right)}}\, \ket \alpha.
\end{equation*}
After measuring the state of the ancillary qubit ($\alpha$), the probability of obtaining the outcome $i$ by performing a subsequent measurement on the index register is given by
\begin{equation*}
\mathbb P(i|\alpha)=\frac{ 1+(-1)^\alpha|\langle\bx|\bx_i\rangle|^2}{ {N+(-1)^\alpha\sum_{i=0}^{N-1} |\langle\bx|\bx_i\rangle|^2}}.
\end{equation*}
As a consequence,
\begin{equation}\label{knn2}
\mathbb Q(i):=\mathbb P(i|0)-\mathbb P(i|1)=\frac{2(|\langle\bx|\bx_i\rangle|^2- C)}{N(1-{C^2})},
\end{equation}
with $C=\frac{1}{N}\sum_{i}|\langle\bx|\bx_i\rangle|^2$ being a constant value. In practice, \eqref{knn2} is proportional to the squared cosine similarity $|\langle\bx_i|\bx\rangle|^2$ between $\bx_i$ and $\bx$. Therefore, by sampling from the index register, it is possible to identify the indexes with the highest $\mathbb Q$ values, i.e., those corresponding to the closest vectors to $\bx$. Actually, since $\mathbb Q$ is proportional to the square of the cosine similarity, the values of each data feature must be concordant in sign in order to extract only the data instances most similar to $\bx$ (and not also the most dissimilar ones).

\subsection{Quantum Binary Classifier}
\label{subsec:quantum-bin-classifier}
\citet{binary_classifier_published} have recently presented a quantum binary classifier based on the cosine similarity metric. Its structure is simple: it iterates the preparation of a superposition state with training and test features encoded as amplitudes, a SWAP test involving states of one qubit, and a final measurement process. Specifically, the measurement outcomes allow estimating a probability value that is directly related to a weighted label assignment with the weights given by the cosine similarity.

More in detail, let $X=\{\bx_i, y_i\}_{i=0,...,N-1}$, with $\bx_{i}\in\mathbb R^d$ and $y_i\in\{-1,1\}$ $\forall i\in\{0,...,N-1\}$, be a training set of $N$ data instances represented in a real feature space of dimension $d$ and characterised by two-valued labels, and let $\bx\in\mathbb R^d$ be a new (test) data instance to be classified as either $-1$ or $1$. Let us take into account the following (classical) classification model:
\begin{equation}\label{model}
y(\bx):=\mbox{sgn}\left(\sum_{i=0}^{N-1} y_i\cos (\bx_i, \bx)\right),
\end{equation}
where $\cos(\bx_i,\bx):=\frac{\bx_i\cdot\bx}{\norm{\bx_i}\norm{\bx}}$ is the cosine similarity between the training vector $\bx_i$ and $\bx$. In this model \eqref{model}, any training vector contributes to the prediction of the new label, and such a contribution is weighted by the cosine similarity with respect to the new instance. Now, let us consider a $\log_2 N$-qubit register to encode the indexes of the training data vectors, a $n$-qubit register (with $n=\log_2 d$) to store the data instances within the amplitude encoding, and a single qubit to encode the values of the labels according to $b_i=\frac{1-y_i}{2}\in\{0,1\}$. Then, let us construct the state
\begin{equation}\label{X}
\ket X=\frac{1}{\sqrt N}\sum_{i=0}^{N-1}\ket i\ket{\bx_i}\ket{b_i}\in \sH_{index}\otimes\sH_n\otimes \sH_l,
\end{equation}
with $\sH_l$ being the Hilbert space of the label qubit. The state in question \eqref{X} encodes the training set $X$ as a quantum superposition of its elements and the respective labels; note that one qubit is sufficient for the encoding of all the labels. In addition, in the same registers, let us construct the state
\begin{equation}\label{psi}
\ket{\psi_\bx}=\frac{1}{\sqrt N}\sum_{i=0}^{N-1}\ket i \ket\bx\ket -\in \sH_{index}\otimes\sH_n\otimes \sH_l,
\end{equation}
where the label qubit is in the state $\ket-=\frac{1}{\sqrt 2}(\ket 0-\ket 1)$; in this way, the test data vector $\bx$ is represented in a quantum superposition of the two possible classes. To allow the coexistence of states \eqref{X} and \eqref{psi} in the same registers, let us consider an ancillary qubit ($a$), and let us prepare the superposition state
\begin{equation}\label{Phi}
\frac{1}{\sqrt 2} \left(\ket X \ket 0 +\ket{\psi_\bx}\ket 1\right)\in\sH_{index}\otimes\sH_n\otimes\sH_l\otimes\sH_a.
\end{equation}
The state \eqref{Phi} is entangled. Indeed, there is a quantum correlation between the state of the ancillary qubit $a$ and the content of the registers. After the preparation of the initial state, let us perform a SWAP test between a second ancillary qubit ($b$), initialized in $\ket +=\frac{1}{\sqrt 2}(\ket 0+\ket 1)$, and the qubit $a$. Let $c$ be the control qubit (initialized in $\ket 0$) in the SWAP test; a straightforward calculation shows that the probability of obtaining the outcome 1 by measuring the qubit $c$ is
\begin{equation*}
\mathbb P(1)=\frac{1}{4}(1-\langle X|\psi_\bx\rangle),
\end{equation*}
which is directly related to the considered classification model (\ref{model}), since 
\begin{equation*}
\langle X|\psi_\bx\rangle=\frac{1}{N\sqrt 2}\sum_{i=0}^{N-1} y_i\cos(\bx_i,\bx).\hspace{0.8cm}
\end{equation*}
Therefore, given the probability $\mathbb P(1)$ or an estimate thereof, it is possible to predict the label of $\bx$ (according to model \ref{model}) by means of
\begin{equation}\label{prediction}
y(\bx)=\mbox{sgn}\left[1-4\,\mathbb P(1)\right].
\end{equation}

\section{Quantum Pipeline}
\label{sec:quantum-pipeline}
This section presents the quantum pipeline that has been implemented and tested, providing information about the components, the implementation, and the complexity. The code is available at \url{https://github.com/ZarHenry96/quantum-ml-pipeline}.

\subsection{Components}
\label{subsec:components}
\begin{figure}[tb]
    \vspace{20pt}
    \centering
    \includegraphics[width=\textwidth]{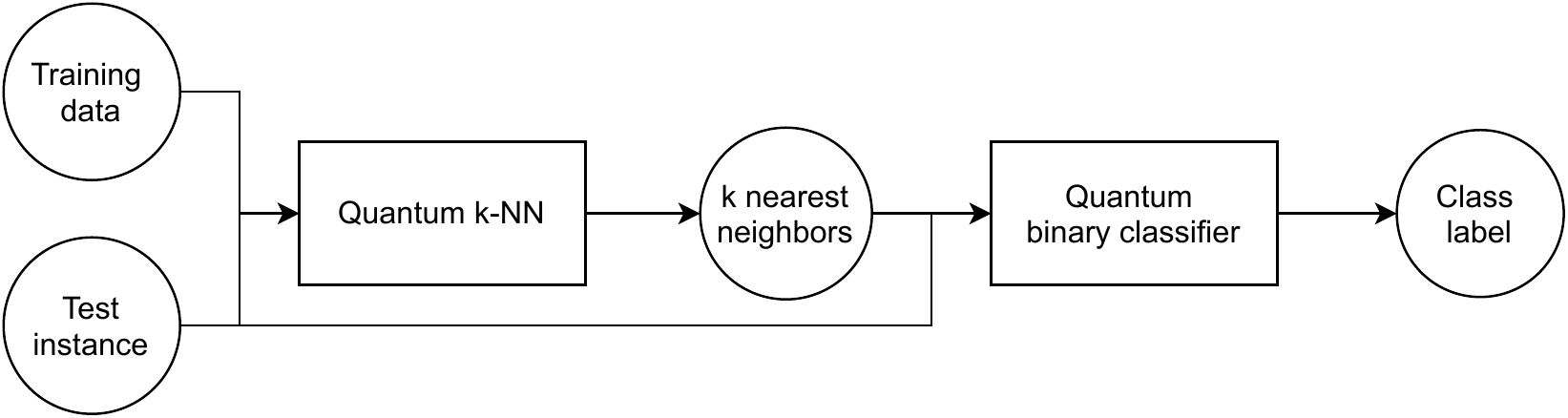}
    \caption{Quantum pipeline workflow overview.}
    \label{fig:q-pipeline-workflow-overview}
    \vspace{20pt}
\end{figure}

The quantum pipeline evaluated in this work consists of a quantum $k$-NN followed by a quantum binary classification model, with the quantum $k$-NN providing the nearest neighbors as input to the subsequent model. The workflow is displayed in \cref{fig:q-pipeline-workflow-overview}.

Concerning the quantum $k$-NN, several variants exist in the literature, as depicted in \cref{subsec:k-nn-n-quantum-k-nns}. However, some aspects must be considered. First, the variants based on the Hamming distance can not be applied directly to the most common problems, namely, the ones characterised by real-valued features, since the considered metric represents a distance on binary strings. Second, most variants include an oracle-based algorithm originating from Grover's, such as D\"{u}rr's or the amplitude estimation. As a consequence, they require a problem-dependent black-box function in order to be used; this negatively affects their ease of use and implementation. In addition, the usage of the amplitude estimation algorithm to transfer distance information to qubits states has, in turn, representation issues in the case of real-valued features: an approximation of the estimated distances is obliged. All these factors make the variant proposed by \citet{qknn_implemented}, described in detail in \cref{subsubsec:quantum-knn-basheer}, the best candidate for experiments on real-valued datasets. Specifically, the parallel processing of multiple test instances suggested by \citet{qknn_cat_tensor_networks} has not been taken into account here.

The quantum binary classification model selected for being combined with the quantum $k$-NN is the classifier described in \cref{subsec:quantum-bin-classifier} \citep{binary_classifier_published}, whose structure is quite simple and very similar to that of the chosen quantum $k$-NN. Actually, the possibility of building a pipeline using these two quantum models was briefly anticipated in the pre-print version of the same article \citep{binary_classifier}.

\subsection{Implementation}
\label{subsec:implementation}
The quantum pipeline has been implemented using Qiskit, i.e., the open-source SDK provided by IBM \citep{qiskit}. Qiskit allows building quantum circuits using the Python programming language, and the circuit execution can be performed either on simulators or real quantum devices. Several simulation backends are provided by IBM, and it is possible not only to get measurement counts as in real quantum devices but also to retrieve the state vector of the circuit at any point of the execution, namely, the amplitude of each state. 

In practice, exploiting Qiskit, code has been developed that automatically initializes and builds circuits for the quantum algorithms involved (quantum $k$-NN, quantum binary classifier) based on the dataset provided as input (i.e., the number of samples, the number of features, and features values). Moreover, the following execution modalities have been implemented for both algorithms: \textit{classical}, which does not build any quantum circuit but runs the corresponding classical algorithm; \textit{statevector}, which processes the final state vector of the circuit to provide the output, thus representing an ideal execution with an infinite number of runs; \textit{simulation} (named \textit{local simulation} in the code), which provides counts by sampling from the final probability distribution; \textit{online simulation}, which is the same as \textit{simulation} but on hardware provided by IBM; and \textit{quantum}, which exploits real IBM quantum devices. Concerning the classical modality, the distance metric used for the $k$-NN is the cosine distance; the reason lies in the fact that the chosen quantum $k$-NN selects the $k$ nearest neighbors based on the fidelity, which corresponds to the squared cosine similarity in the case of pure quantum states (see Equation \ref{eq:fidelity}). Instead, for the binary classifier, \cref{model} is used. It is also worth highlighting that no noise has been taken into account in the execution of gates in any simulated modality (including the \textit{statevector}), and that the execution modality does not need to be the same for the pipeline components. Eventually, the possibility of retrieving the results of online executions at a later moment has been implemented due to the presence of long waiting times for the quantum devices. 

\begin{algorithm}[t!]
    \SetNoFillComment

    \KwIn{training data $D$, test instance $x$, number of nearest neighbors $k$, execution modalities (not \textit{classical}) for the two components $exec\_mods$}
    \KwResult{class label $label$ $\in \{-1,1\}$}
    
    $D,\ x \gets normalization(D,\ x)$\; 
    \BlankLine
    
    \tcc{Quantum $k$-NN}
    $circ_{qknn} \gets buildQKNNCircuit(D,\ x,\ exec\_mods[0])$\tcp*{See \cref{fig:qknn-circ}}
    $res_{qknn} \gets execute(circ_{qknn},\ exec\_mods[0])$\;
    $k\_nn \gets getKNearestNeighbors(D,\ k,\ res_{qknn},\ exec\_mods[0])$\;
    \BlankLine
    
    \tcc{Quantum binary classifier}
    $circ_{qbc} \gets buildQBCCircuit(k\_nn,\ x,\ exec\_mods[1])$\tcp*{See \cref{fig:qbc-circ}}
    $res_{qbc} \gets execute(circ_{qbc},\ exec\_mods[1])$\;
    $label \gets getLabel(res_{qbc},\ exec\_mods[1])$\;
    \BlankLine
    
    \Return{$label$}\;
    
    \caption{Quantum Pipeline.}
    \label{alg:q-pipeline}
\end{algorithm}

\begin{figure}[b!]
    \centering
    \begin{subfigure}{0.96\textwidth}
        \centering
        \includegraphics[trim={0 1.8cm 0 0}, clip, width=\textwidth]{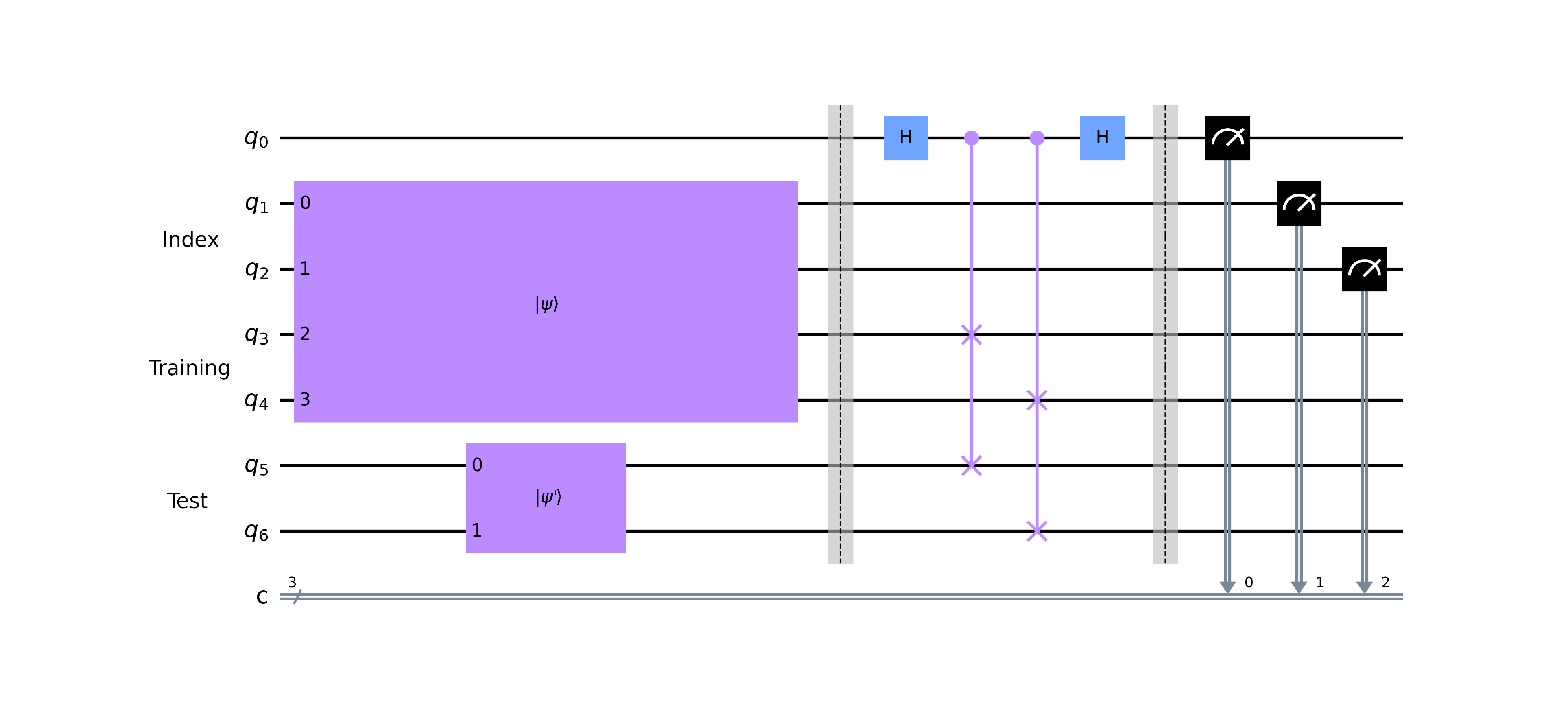}
        \caption{Quantum $k$-NN.}
        \label{fig:qknn-circ}
    \end{subfigure} \\ \vspace{12pt}
    \begin{subfigure}{0.96\textwidth}
        \centering
        \includegraphics[width=\textwidth]{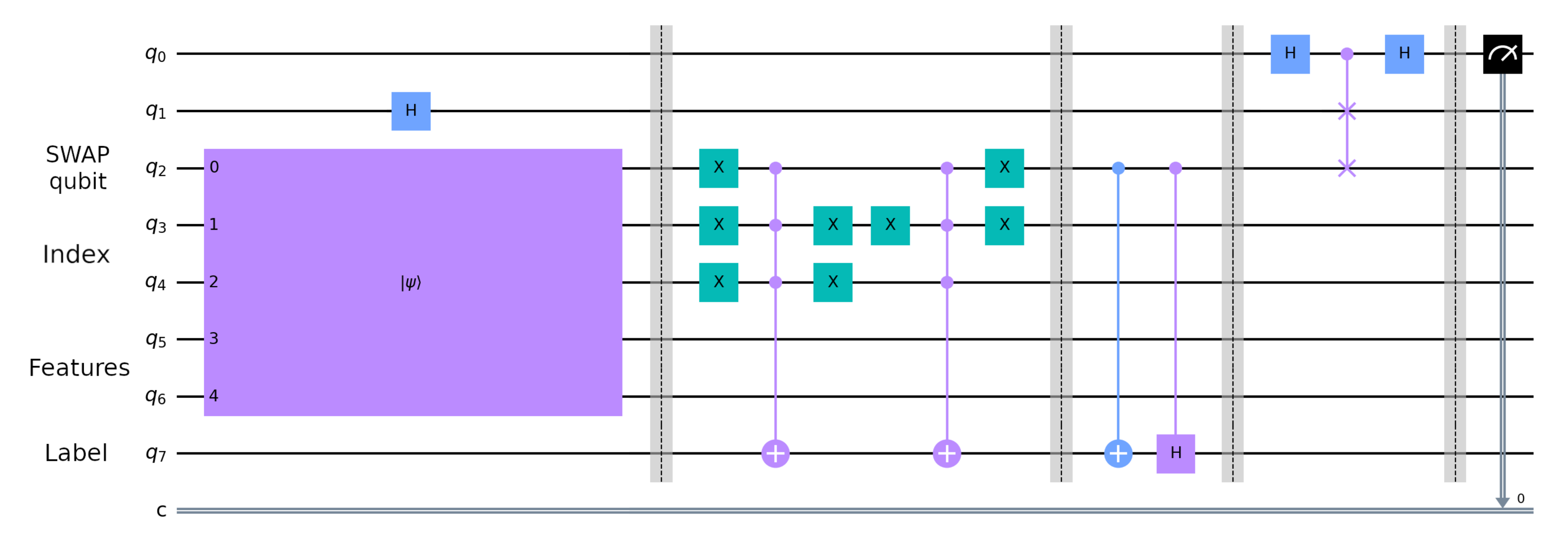}
        \caption{Quantum binary classifier.}
        \label{fig:qbc-circ}
    \end{subfigure} \\
    \caption{Quantum pipeline circuits example. The first one (a) corresponds to the quantum $k$-NN, the second one (b) to the quantum binary classifier. In the case of the \textit{statevector} modality, the final measurements are not present. The barriers (vertical dotted lines) have been added for illustrative purposes.}
    \label{fig:quantum-circuits}
\end{figure}

The pseudocode of the quantum pipeline is shown in \cref{alg:q-pipeline}. Actually, the pseudocode in question is valid for all execution modalities but the \textit{classical}, which does not require circuits. If the \textit{classical} modality is selected for one component, the corresponding block is replaced with the execution of the classical version of the algorithm. In any case, the first step is the unit-norm normalization of the training and test data (Line 1). This operation is essential for the amplitude encoding of data and consists in dividing the features of each instance by the instance norm. If a data instance has the norm equal to zero, all its attributes are replaced with an epsilon value ($0.000001$) to allow the normalization. The procedure in question is executed also in the case of a quantum pipeline with only classical components. Moreover, in the experiments, an additional normalization procedure has been applied to the data before the unit-norm one as part of the experimental setup common to all methods tested; the details are provided in \cref{subsec:exp-setup}. It is also worth mentioning that the values of each data feature must be concordant in sign, for the quantum $k$-NN to work properly (in the experiments, this has been ensured by the additional normalization procedure just mentioned); furthermore, the quantum binary classifier requires the training data class labels to be in $\{-1,1\}$.

Focusing on the non-classical modalities, the step following the data normalization is the construction of the quantum $k$-NN circuit based on the normalized data (Line 2). A sample circuit is shown in \cref{fig:qknn-circ}. In detail, a quantum $k$-NN circuit consists of three main slices (the vertical barriers have been added for illustrative purposes): registers initialization, SWAP test without measurement, and final measurements. In the first slice, the training data index is set up, simultaneously to the encoding of the training and the test features. Specifically, the index ($q_1$-$q_2$) and the training features ($q_3$-$q_4$) registers are jointly initialized for simplicity; indeed, they are entangled according to the quantum $k$-NN algorithm. After that, the SWAP test without measurement is performed by means of two Hadamard gates and the controlled SWAP gates, whose number increases linearly with the number of feature qubits. In particular, the SWAP gates act on the training ($q_3$-$q_4$) and the test features ($q_5$-$q_6$) registers. Eventually, the state of the first ancillary qubit (the SWAP test measurement qubit $q_0$) and of the index register is measured; obviously, in the case of the \textit{statevector} modality, the measurements are not present. Regarding the number of qubits required ($qubits_{qknn}$), which is dataset dependent, it is given by
\begin{equation}
    \label{eq:qknn-qubits-num}
    qubits_{qknn} = 1 + qubits_{qknn\_index} + 2 * qubits_{features}\,,
\end{equation}
where the value $1$ corresponds to the SWAP test measurement qubit ($q_0$), $qubits_{qknn\_index}$ represents the number of index qubits ($q_3$-$q_4$), and $qubits_{features}$ is the number of feature qubits (either $q_3$-$q_4$ or $q_5$-$q_6$). 

Once the quantum $k$-NN circuit has been built, it is executed according to the specified modality (Line 3). Actually, if the circuit includes measurements, it is executed multiple times to provide the desired number of counts ($simulation\_shots$ parameter described in \cref{subsec:exp-setup}). Finally, the execution output ($res_{qknn}$), which corresponds to either a state vector or state counts (the number of times each state has been observed), is processed in order to extract the $k$ nearest neighbors (Line 4). In detail, the amplitudes/counts are exploited to estimate the quantity $P(i|0)-P(i|1)$, with $i$ being a training data index and the $0$/$1$ value being the state of $q_0$ (see Equation \ref{knn2}). Then, the training data are sorted according to this quantity, which is proportional to the similarity with respect to the test instance, allowing the identification of the nearest neighbors.

The next step is the construction of the quantum binary classifier circuit based on the selected $k$ nearest neighbors and the normalized test instance (Line 5). An example circuit is displayed in \cref{fig:qbc-circ}. In particular, it is possible to distinguish five main slices in the circuit (also in this case, the vertical barriers have been added for illustrative purposes): registers initialization, training data labels configuration, test label set up, SWAP test without measurement, and final measurement. Concerning the initialization, the first qubit subjected to the swap ($q_1$) is set to the uniform superposition of $0$ and $1$ ($\ket{+}$). Instead, the second one ($q_2$) is entangled with the register ($q_3$-$q_4$) representing the training data index and the register ($q_5$-$q_6$) in which both the training and the test features are encoded in superposition. Hence, it is jointly initialized with all them (from $q_3$ to $q_6$) for simplicity. Regarding the qubit representing the label ($q_7$), it takes only well defined values ($0$, $1$, and the uniform superposition of them), thus it is configured separately (second and third slices of the circuit) to limit the complexity of the joint initialization operation. More in detail, the training data labels (second slice) are encoded in the last qubit of the circuit by selecting the desired index register states through NOT gates (denoted as $X$s in the image) and setting the corresponding label states through multi-controlled NOT gates (this procedure is required only for one label value, the one associated with the $1$ state, namely, $-1$). The NOT gate applied to the last SWAP qubit ($q_2$) at the beginning and at the end of this step is necessary in order to work on the states associated with the training data. Instead, for the test instance (third slice of the circuit), the label qubit is set to the $\ket{-}$ state according to the algorithm by means of a controlled NOT and a controlled Hadamard gates. Then, the SWAP test without measurement is performed (fourth slice), and the state of the first qubit ($q_0$) is measured (last slice). Also in this case, there is no final measurement for the \textit{statevector} modality. The number of qubits required by the quantum binary classifier for a generic dataset ($qubits_{qbc}$) is given by
\begin{equation}
    \label{eq:qbc-qubits-num}
    qubits_{qbc} = 3 + qubits_{qbc\_index} + qubits_{features} + 1\,.
\end{equation}
In detail, the value $3$ corresponds to the qubits needed by the SWAP test ($q_0$-$q_1$-$q_2$), $qubits_{qbc\_index}$ represents the number of index qubits ($q_3$-$q_4$), $qubits_{features}$ is the number of feature qubits ($q_5$-$q_6$), and the $1$ represents the qubit used for the label encoding ($q_7$).  

After the construction of the quantum binary classifier circuit, there is the execution step (Line 6); the considerations about the presence of measurements made for the quantum $k$-NN also hold for the classifier. Eventually, the execution output is processed (Line 7) in order to predict the test instance label. In particular, the amplitudes/counts are used to estimate the probability $P(1)$ of obtaining $1$ by measuring the state of the qubit $q_0$. The probability in question allows predicting the class label ($-1$ if $P(1) > 0.25$, $1$ otherwise, according to Equation \ref{prediction}), which is then returned as the last operation of the pipeline (Line 8).

\subsection{Complexity Observations}
\label{subsec:complexity}
\citet{qknn_implemented} define the gate complexity of their quantum $k$-NN algorithm, which is equal to $O(log_2\,d)$, with $d$ being the number of data features. However, the complexity in question is expressed in terms of controlled SWAP (Fredkin) gates, which are not elementary gates. Moreover, the registers initialization is not included because they assume the presence of an initialization quantum oracle. Instead, for their quantum binary classifier, \citet{binary_classifier_published} provide the overall time complexity, which is equal to $O(\epsilon^{-2}\,log_2(Nd))$, with $\epsilon$ being the desired upper bound to the prediction error, and $N$ being the training dataset size. Nevertheless, the authors assume the presence of a QRAM, i.e., a quantum random access memory \citep{PhysRevLett.100.160501}, from which to retrieve the state given as input to the SWAP test (fourth and fifth slices of the circuit shown in \cref{fig:qbc-circ}).

Concerning the pipeline implementation presented in this work, some observations can be made for execution modalities different from \textit{classical}. First of all, the initial unit-norm normalization step has $O(Nd)$ complexity, since it is necessary to scan all the data features. The construction of the quantum $k$-NN circuit has $O(2^{\lceil log_2\,N\rceil+\lceil log_2\,d\rceil}+\lceil log_2\,d\rceil)$ complexity, with the first term given by the joint index-training initialization and the second one given by the SWAP test. Instead, the complexity of the circuit execution depends on the execution modality and its implementation inside Qiskit; if it is not a \textit{statevector} execution, the complexity is influenced also by the number of measurements. Regarding the number of gates, it is worth highlighting that the registers initialization is an expensive operation. Indeed, the generation of an arbitrary target state requires to find the correct sequence of elementary gates. The controlled SWAP gates also have a significant impact since, as mentioned previously, their number depends on the number of data features and they are not elementary operations. The last quantum $k$-NN step is the extraction of the nearest neighbors, which includes the processing of the execution output. In detail, for a \textit{statevector} execution, it is necessary to process the final state vector of the circuit, with a $O(2^{1+\lceil log_2\,N\rceil+2\lceil log_2\,d\rceil})$ complexity. Instead, for all the other execution modalities, the state counts must be processed, and the corresponding complexity is $O(2^{1+\lceil log_2\,N\rceil})$. In any case, after the execution output processing, the $k$ nearest neighbors are extracted by sorting the index values with a $O(N log_2\,N)$ complexity.

Looking at the second half of the pipeline, namely, the quantum binary classifier, its complexity is related to the number of nearest neighbors $k$. In particular, the construction of the classifier circuit has $O(2^{1+\lceil log_2\,k\rceil+\lceil log_2\,d\rceil}+k\,2\lceil log_2\,k\rceil)$ complexity, with the second term being a worst case estimate for the training labels set up. Regarding the circuit execution and the number of gates, the observations made for the quantum $k$-NN hold also for the classifier. Actually, there is only one controlled SWAP gate in this case. Nevertheless, there could be several (maximum $k$) multi-controlled CNOT gates, which in turn have a significant impact on the performance. Eventually, there is the execution output processing, which has $O(2^{4+\lceil log_2\,k\rceil+\lceil log_2\,d\rceil})$ complexity if the execution modality is \textit{statevector}, $O(1)$ complexity otherwise. The final label prediction has constant complexity.

To give an idea of the runtimes, for $N=168$, $d=12$, and $k=9$, the pipeline execution time on the machine used in the experiments (whose specifications are provided at the beginning of \cref{sec:empirical-eval}) is in the order of 2-3 seconds. This holds for both the \textit{statevector} and the \textit{simulation} modalities, with both components having the same execution modality. Actually, among the two, the \textit{simulation} modality turns out to be a little more time consuming. In the situation just described, the circuit size is $17$ qubits for the quantum $k$-NN and $12$ qubits for the quantum binary classifier.

\section{Empirical Evaluation}
\label{sec:empirical-eval}
This section deals with the quantum and classical algorithms taken into account, the datasets selection and preparation, the setup of the experiments, and the results obtained. In particular, the experiments have been executed on a shared machine with an Intel Xeon Gold 6238R processor running at 2.20GHz and 125 GB of RAM.

\begin{table}[t!]
    \centering
    \begin{subtable}[t]{.328\textwidth}
        \centering
        \begin{tabular}{c}
            \textbf{Quantum pipeline} \\ \hline
            classical - classical \\ \hline
            statevector - classical \\ \hline
            classical - statevector \\ \hline
            statevector - statevector \\ \hline
            simulation - classical \\ \hline
            classical - simulation \\ \hline
            simulation - simulation \\
        \end{tabular}
        \caption{}
        \label{tab:pipelines}
    \end{subtable}
    \begin{subtable}[t]{.328\textwidth}
        \centering
        \begin{tabular}{c}
            \textbf{Quantum classifier} \\ \hline
            statevector \\ \hline
            simulation \\ \arrayrulecolor{white}\hline
             \\ \arrayrulecolor{white}\hline
             \\ \arrayrulecolor{white}\hline
             \\ \arrayrulecolor{white}\hline
             \\ \arrayrulecolor{white}\hline
             \\
        \end{tabular}
        \caption{}
        \label{tab:q-classifier}
    \end{subtable}
    \begin{subtable}[t]{0.328\linewidth}
        \centering
        \begin{tabular}{c}
            \textbf{Baseline method} \\ \hline
            random forest \\ \hline
            SVM \\ \hline
            $k$-NN \\ \hline
            $k$-NN + classifier \\ \hline
            $k$-NN + SVM \\ \arrayrulecolor{white}\hline
             \\ \arrayrulecolor{white}\hline
             \\
        \end{tabular}
        \caption{}
        \label{tab:baselines}
    \end{subtable}
    \caption{Quantum pipeline modalities (a), quantum binary classifier modalities (b), and baseline methods (c) considered.}
    \label{tab:methods}
    \vspace{-12pt}
\end{table}

\subsection{Methods}
\label{subsec:methods}
The "quantum $k$-NN" \citep{qknn_implemented} - "quantum binary classifier" \citep{binary_classifier_published} pipeline described in \cref{sec:quantum-pipeline} has been tested under different execution modality combinations, which are reported in \cref{tab:pipelines}. In detail, these experiments aimed at understanding the performance of the classical pipeline, verifying the equivalence between \textit{classical} and \textit{statevector} (the latter represents an ideal execution), and analysing the impact of simulating the pipeline components. Initially, pipelines including quantum executions were planned too. Nevertheless, the quantum device of interest was dismissed before the experiments could have been executed and no equivalent machine (in the number of qubits) has been made available yet (at time of writing, the maximum number of qubits available for a free account is five).

The quantum binary classifier alone has been evaluated using the modalities reported in \cref{tab:q-classifier}. In particular, the \textit{classical} modality has not been taken into account due to the effective equivalence with respect to \textit{statevector}. The purpose of these experiments was to collect data to confirm a potential improvement in the performance of the model with the introduction of locality in the form of the quantum $k$-NN.

Eventually, the pipeline performance have been compared with some classical baseline methods, which are reported in \cref{tab:baselines}. Most of these algorithms have been taken directly from scikit-learn \citep{scikit-learn}, but others have been composed starting from scikit-learn procedures. In detail, the default parameters provided by scikit-learn have been used for the random forest (e.g., the number of trees has been set to 100), whereas, for the SVM, two different kernels have been tested, i.e., Gaussian and linear. Two distance metrics have also been evaluated for the $k$-NN: the cosine and the Euclidean distance. Finally, two model pipelines have been considered. The \textit{k-NN + classifier} represents the classical analogue of the implemented quantum pipeline. Indeed, the classifier in question is the binary classifier based on the cosine similarity defined in \cref{model}. Instead, the \textit{k-NN + SVM} has been taken into account in order to evaluate the benefits of replacing the binary classifier with a more complex model like the SVM. Actually, these baseline pipelines have been tested using both metrics (cosine and Euclidean) for the $k$-NN and both kernels (Gaussian and linear) for the SVM. It is also worth mentioning that the \textit{k-NN + classifier} with cosine distance metric differs from the \textit{classical - classical} execution of the quantum pipeline for the absence of the input data unit-norm normalization.

\subsection{Datasets}
\label{subsec:datasets}
All the datasets used in the experiments are from the UCI Machine Learning Repository \citep{uci_ml_repo}. In detail, they have been selected according to the following criteria: the associated machine learning task is classification (all the considered models are classifiers); the attributes are real-valued, preferably, but some integer features are accepted (actually, the \textit{02\_transfusion} dataset has only integer attributes but is marked as real-valued on the UCI site); the number of features is less than or equal to 16; the number of instances is reasonable, namely, less than one thousand.

The reason for the filter on the type of attributes lies in the amplitude encoding exploited by the considered quantum models. In particular, numerical data are required, and integer values are acceptable because of the unit-norm normalization. Regarding the limit on the number of features, it is related to the experiments originally planned on the real quantum device. The number of qubits of that machine was 15. Hence, if the number of qubits required to encode the features ($qubits_{features}$) had exceeded four (more than 16 attributes), the number of embeddable training instances in the quantum $k$-NN circuit would have been less than 17 ($qubits_{qknn\_index} \leq 4$), a too-small quantity (see Equation~\ref{eq:qknn-qubits-num}). Eventually, the constraint on the number of instances was aimed at allowing the encoding of the dataset in the circuit without the need for a too drastic subsampling.

\begin{table}[b!]
    \hypersetup{hidelinks}
    \centering
    \begin{tabular}{c|c|c|c|c|c}
        \multicolumn{1}{c|}{\multirow{2}{*}{\textbf{Name}}} & \textbf{Original} & \textbf{Original} & \textbf{Features} & \textbf{Size} & \textbf{Size} \\ 
         & \textbf{size} & \textbf{classes \#} & \textbf{\#} & \textbf{(15 qb.)} & \textbf{(32 qb.)}  \\ \hline
        \href{https://archive.ics.uci.edu/ml/datasets/Iris}{01\_iris\_setosa\_versicolor} & \multicolumn{1}{c|}{\multirow{3}{*}{150}} & \multicolumn{1}{c|}{\multirow{3}{*}{3}} & \multicolumn{1}{c|}{\multirow{3}{*}{4}} & 100 & - \\ \cline{1-1} \cline{5-6}
        \href{https://archive.ics.uci.edu/ml/datasets/Iris}{01\_iris\_setosa\_virginica} & & & & 100 & - \\ \cline{1-1} \cline{5-6}
        \href{https://archive.ics.uci.edu/ml/datasets/Iris}{01\_iris\_versicolor\_virginica} & & & & 100 & - \\ \hline
        \href{https://archive.ics.uci.edu/ml/datasets/Blood+Transfusion+Service+Center}{02\_transfusion} & \multicolumn{1}{c|}{\multirow{2}{*}{748}} & \multicolumn{1}{c|}{\multirow{2}{*}{2}} & \multicolumn{1}{c|}{\multirow{2}{*}{4}} & \multicolumn{1}{c|}{\multirow{2}{*}{748}} & \multicolumn{1}{c}{\multirow{2}{*}{-}} \\
        \citep{transfusion} & & & & & \\ \hline
        \href{https://archive.ics.uci.edu/ml/datasets/Vertebral+Column}{03\_vertebral\_column\_2C} & 310 & 2 & 6 & 310 & - \\ \hline
        \href{https://archive.ics.uci.edu/ml/datasets/seeds}{04\_seeds\_1\_2} & 210 & 3 & 7 & 140 & - \\ \hline
        \href{https://archive.ics.uci.edu/ml/datasets/Ecoli}{05\_ecoli\_cp\_im} & 336 & 8 & 7 & 220 & - \\ \hline
        \href{https://archive.ics.uci.edu/ml/datasets/Glass+Identification}{06\_glasses\_1\_2} & 214 & 6 & 9 & 80 & 146 \\ \hline
        \href{https://archive.ics.uci.edu/ml/datasets/Breast+Tissue}{07\_breast\_tissue\_adi-} & \multicolumn{1}{c|}{\multirow{2}{*}{106}} & \multicolumn{1}{c|}{\multirow{2}{*}{6}} & \multicolumn{1}{c|}{\multirow{2}{*}{9}} & \multicolumn{1}{c|}{\multirow{2}{*}{71}} & \multicolumn{1}{c}{\multirow{2}{*}{-}} \\
        \href{https://archive.ics.uci.edu/ml/datasets/Breast+Tissue}{\_fadmasgla} & & & & & \\ \hline
        \href{https://archive.ics.uci.edu/ml/datasets/Breast+Cancer+Coimbra}{08\_breast\_cancer} & \multicolumn{1}{c|}{\multirow{2}{*}{116}} & \multicolumn{1}{c|}{\multirow{2}{*}{2}} & \multicolumn{1}{c|}{\multirow{2}{*}{9}} & \multicolumn{1}{c|}{\multirow{2}{*}{80}} & \multicolumn{1}{c}{\multirow{2}{*}{116}} \\ 
        \citep{breast_cancer_coimbra} & & & & & \\ \hline
        \href{https://archive.ics.uci.edu/ml/datasets/Speaker+Accent+Recognition}{09\_accent\_recognition-} & \multicolumn{1}{c|}{\multirow{2}{*}{329}} & \multicolumn{1}{c|}{\multirow{2}{*}{6}} & \multicolumn{1}{c|}{\multirow{2}{*}{12}} & \multicolumn{1}{c|}{\multirow{2}{*}{80}} & \multicolumn{1}{c}{\multirow{2}{*}{210}} \\ 
        \href{https://archive.ics.uci.edu/ml/datasets/Speaker+Accent+Recognition}{\_uk\_us} & & & & & \\ \hline
        \href{https://archive.ics.uci.edu/ml/datasets/Leaf}{10\_leaf\_11\_9} & \multicolumn{1}{c|}{\multirow{2}{*}{340}} & \multicolumn{1}{c|}{\multirow{2}{*}{30}} & \multicolumn{1}{c|}{\multirow{2}{*}{14}} & \multicolumn{1}{c|}{\multirow{2}{*}{30}} & \multicolumn{1}{c}{\multirow{2}{*}{-}} \\
        \citep{leaf} & & & & & \\
    \end{tabular}
    \caption{Datasets properties (the dataset names are links that lead to the corresponding UCI pages). Note: "qb." stands for qubits.}
    \label{tab:datasets}
\end{table}

The well-formed datasets matching the just presented criteria were 10; indeed, some have been discarded due to missing fields or unclear structure. However, most of these datasets were characterised by more than two classes, whereas Pastorello and Blanzieri's classifier (and also the classical SVM) works on binary labels. Therefore, only the two most represented classes have been retained, mapping them to $\{-1, 1\}$ and discarding the other instances (in the case of a tie, the first two classes have been chosen). Moreover, if the consequent dataset size was still exceeding the number of instances encodable in the quantum $k$-NN circuit using 15 qubits, a random subsampling maintaining the ratio between classes has been applied. Concerning this last step, the size reduction due to the split in training and test set, which is accurately described in the next section, has also been taken into account. The resulting datasets and their properties are reported in \cref{tab:datasets}. 

It is worth highlighting some aspects about the data shown in the table: the dataset name includes a suffix representing the names of the selected classes if they originally were more than two; the Iris dataset ($01$) represents an exception since all the possible combinations between classes have been taken into account, leading the total number of datasets to 12; the suggested three classes merging has been applied to \textit{07\_breast\_tissue\_adi\_fadmasgla}. Regarding the sizes of the datasets used in the experiments, they are reported in the last two columns: the former (\textit{Size, 15 qb.}) contains the sizes resulting after performing the entire process described in the previous paragraph, whereas the latter (\textit{Size, 32 qb.}) shows the sizes without the final subsampling step for the datasets for which was needed. In particular, the three datasets that required the subsampling (i.e., \textit{06\_glasses\_1\_2}, \textit{08\_breast\_cancer}, and \textit{09\_accent\_recognition\_uk\_us}) have been used to analyse the effect of a larger training set, and the value 32 represents the limit number of qubits for an online simulation (far fewer qubits are required in practice).

Another thing that is worth mentioning is the fact that, before reducing its number of classes, the Iris dataset (01) has been modified by correcting the two wrong instances as reported in the Iris UCI page. All the datasets used in the experiments are available together with the code at \url{https://github.com/ZarHenry96/quantum-ml-pipeline}.

\subsection{Experimental Setup}
\label{subsec:exp-setup}
Each method presented in \cref{subsec:methods} has been applied to each dataset reported in \cref{tab:datasets} (both 15 and 32 qubits limit), with the exception of the "quantum binary classifier" - "\textit{02\_transfusion} dataset" pair, since an additional qubit would have been necessary (see Equation~\ref{eq:qbc-qubits-num}). The model evaluation technique chosen is the $k$-fold cross-validation, which works as follows: the dataset is split into $k$ subsets, also called folds; then, $k-1$ folds form the training set, whereas the remainder becomes the test set; the last step is iterated $k$ times so that each fold is used once as the test set. In particular, the stratified $k$-fold cross-validation has been exploited. This means that the folds have been generated in such a way that the ratio between classes in the test set remained as close as possible to that of the original dataset. Eventually, it is worth mentioning that, in all experiments, the same seed has been used for the generation of folds. In this way, all methods have processed exactly the same folds.

\begin{table}[b!]
    \centering
    \begin{tabular}{c|c}
        \textbf{Parameter} & \textbf{Value(s)} \\ \hline
        k\_folds & 5 \\ \hline
        k & 3, 5, 7, 9 \\ \hline
        simulation\_shots & 1024 \\ \hline
        simulation\_runs & 5 \\
    \end{tabular}
    \caption{Parameters of the experiments.}
    \label{tab:exp-params}
\end{table} 

The parameters of the experiments are reported in \cref{tab:exp-params}. In detail, $k\_folds$ represents the number of folds, which has been set to 5. Instead, $k$ corresponds to the number of nearest neighbors selected, a fundamental parameter for both the quantum pipelines, the classical $k$-NN, and the baseline pipelines (\textit{k-NN + classifier} and \textit{k-NN + SVM}). Therefore, several values have been evaluated. Concerning $simulation\_shots$, it represents the number of measurements (hence, circuit executions) performed in the \textit{simulation} modality, and its value is the same for both the quantum $k$-NN and the quantum binary classifier. Finally, $simulation\_runs$ is the number of runs that have been executed for the experiments including either a quantum model ($k$-NN/classifier) in the \textit{simulation} modality or the random forest. Indeed, the methods in question are stochastic, and no seed has been set for them.

In each cross-validation step of each experiment, a min-max data normalization procedure has been applied before executing the model/pipeline. In particular, each attribute in the training set has been rescaled to the interval [0, 1] by subtracting the minimum value and dividing by the range. In the case of a zero range (constant attribute), the attribute has been set to zero. As usual, the test instances have been normalized using the training set parameters (minimum and range values). If a test instance attribute exceeded the interval edges after the normalization (since it was larger/smaller than the maximum/minimum in the training set), it has been clipped to the exceeded edge value. In addition, after the feature normalization, the input data to the quantum models (including the quantum classifier alone) have undergone a unit-norm normalization procedure, as described in \cref{subsec:implementation}. Actually, the min-max normalization has also avoided any sign-related issue for the quantum $k$-NN, which requires the values of each attribute to be concordant in sign.

\subsection{Results}
\label{subsec:results}
The results are presented by means of accuracy-based scatter plots, with the accuracy being defined for a fold as 
\begin{equation*}
    \label{eq:accuracy}
    accuracy = \frac{number\ of\ correctly\ classified\ instances \ in\ the\ fold}{total\ number\ of\ instances\ in\ the\ fold}\,.
\end{equation*}
In the case of multiple runs, the fold average accuracy is reported. Moreover, the statistical significance of the results is verified through the Wilcoxon signed-rank test \citep{wilcoxon_test}, since the data considered are paired. 

In detail, the different execution modalities of both the quantum pipeline and the quantum binary classifier are analysed first. Then, the two quantum models are compared with each other, and the effect of the dataset size on their performance is shown. Finally, the baseline methods are taken into account: an evaluation of the considered distance metrics (cosine/Euclidean) is presented, followed by a comparison of the quantum pipeline and the baseline methods.

\begin{figure}[p]
    \centering
    \begin{subfigure}{0.45\textwidth}
      \centering
      \includegraphics[width=\textwidth]{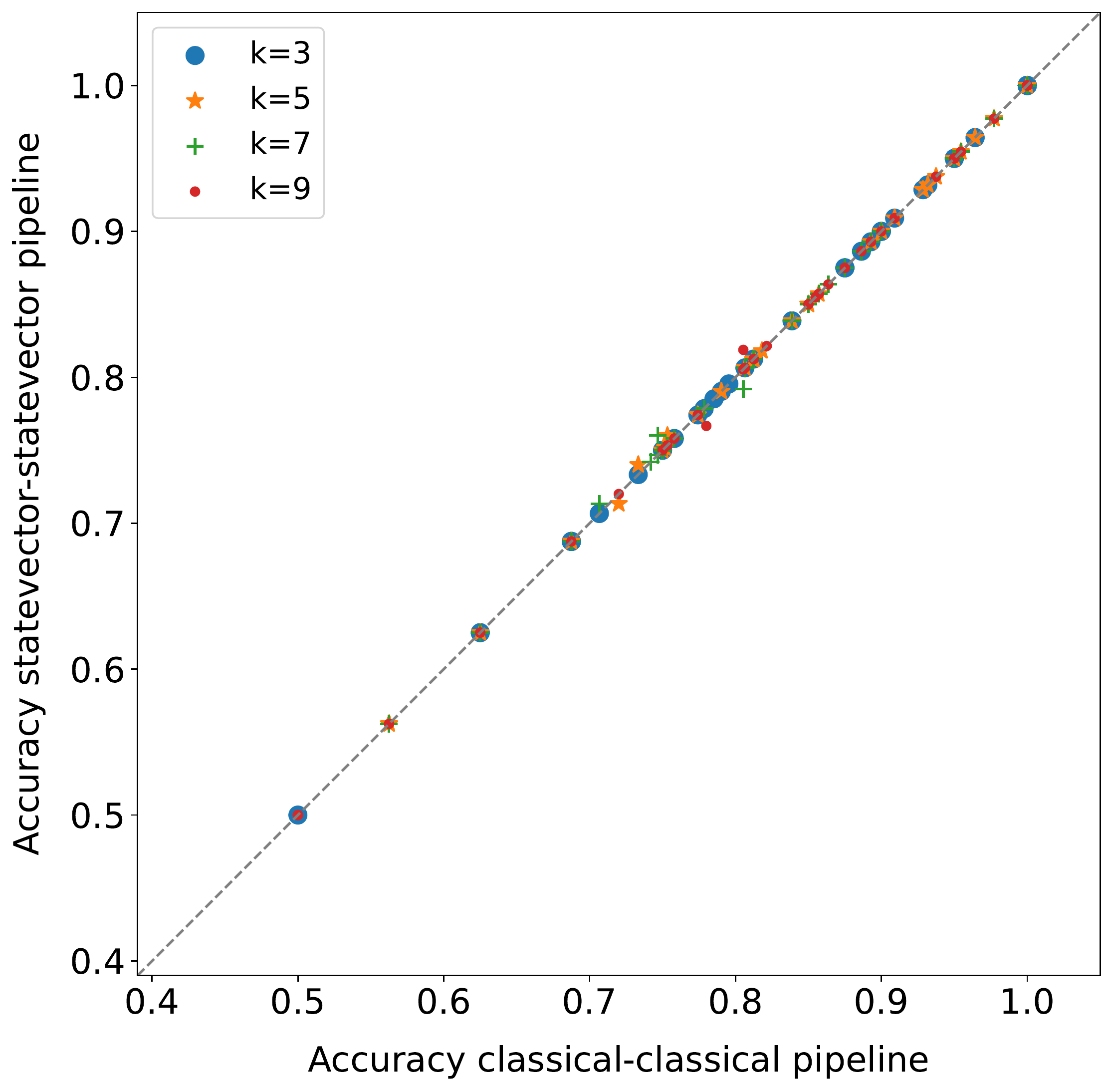}
      \caption{}
      \label{fig:cl-cl-vs-sv-sv}
    \end{subfigure}
    \begin{subfigure}{0.45\textwidth}
      \centering
      \includegraphics[width=\textwidth]{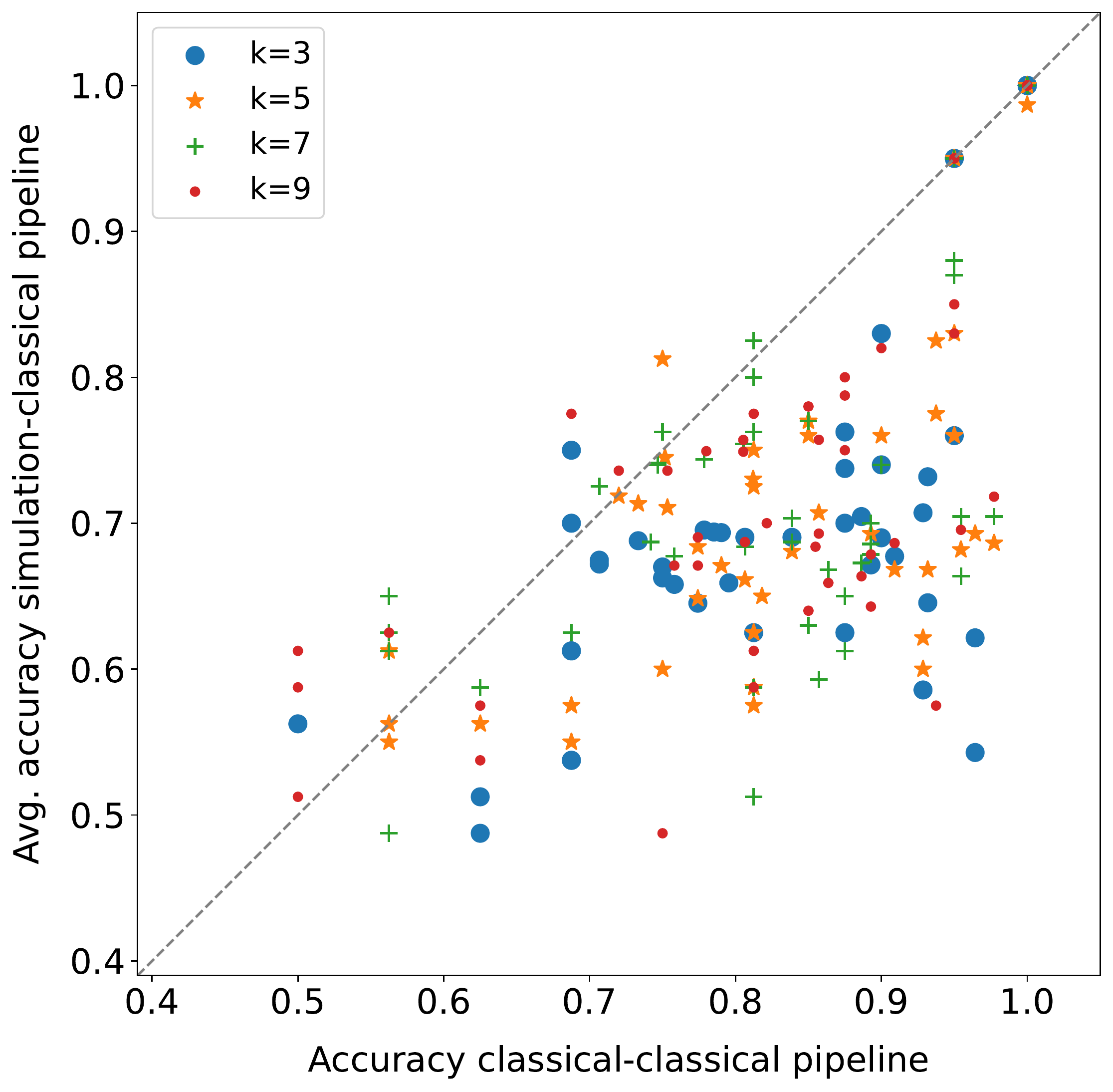}
      \caption{}
      \label{fig:cl-cl-vs-si-cl}
    \end{subfigure} \\ \vspace{8pt}
    \begin{subfigure}{0.45\textwidth}
      \centering
      \includegraphics[width=\textwidth]{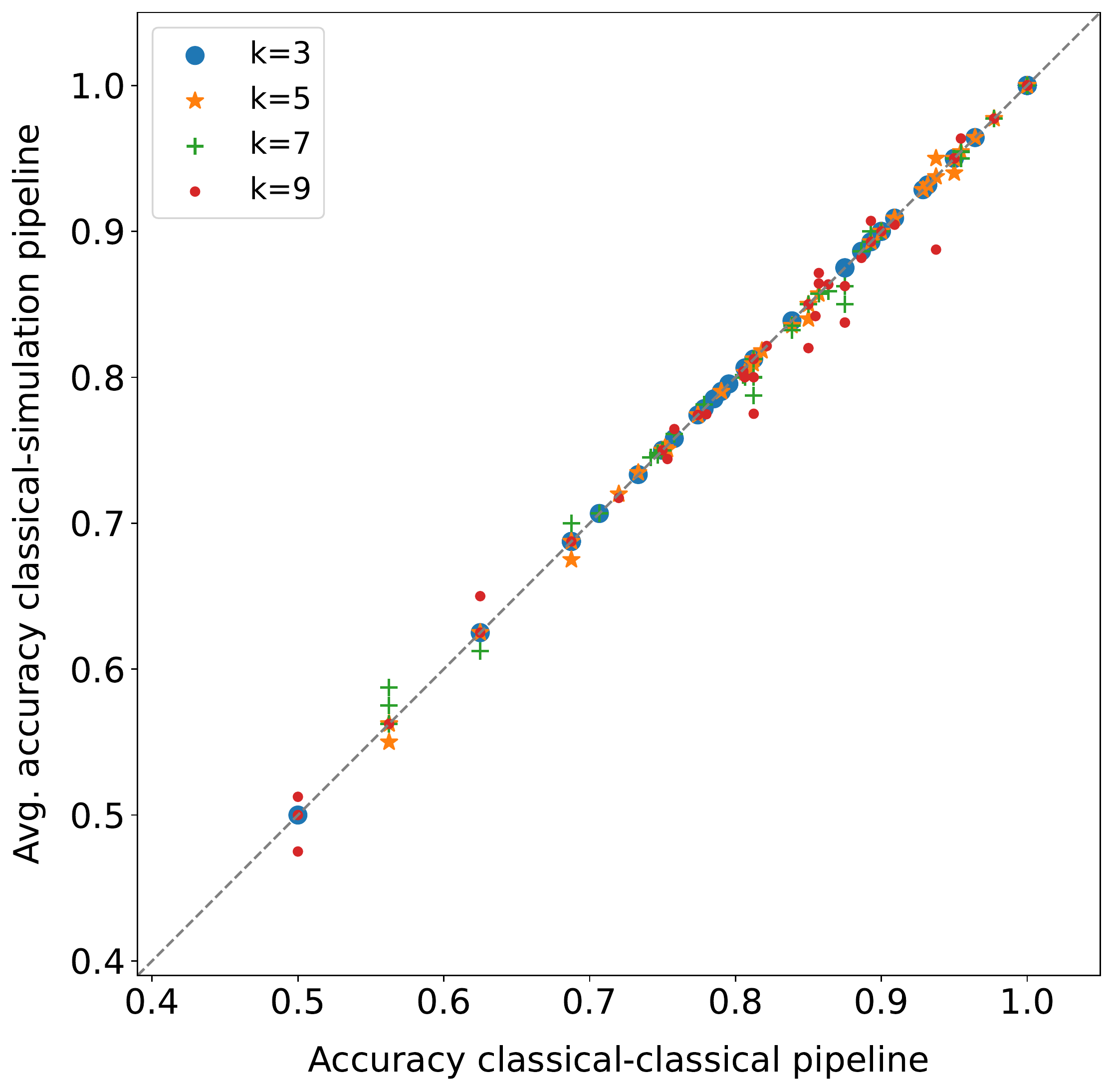}
      \caption{}
      \label{fig:cl-cl-vs-cl-si}
    \end{subfigure} 
    \begin{subfigure}{0.45\textwidth}
      \centering
      \includegraphics[width=\textwidth]{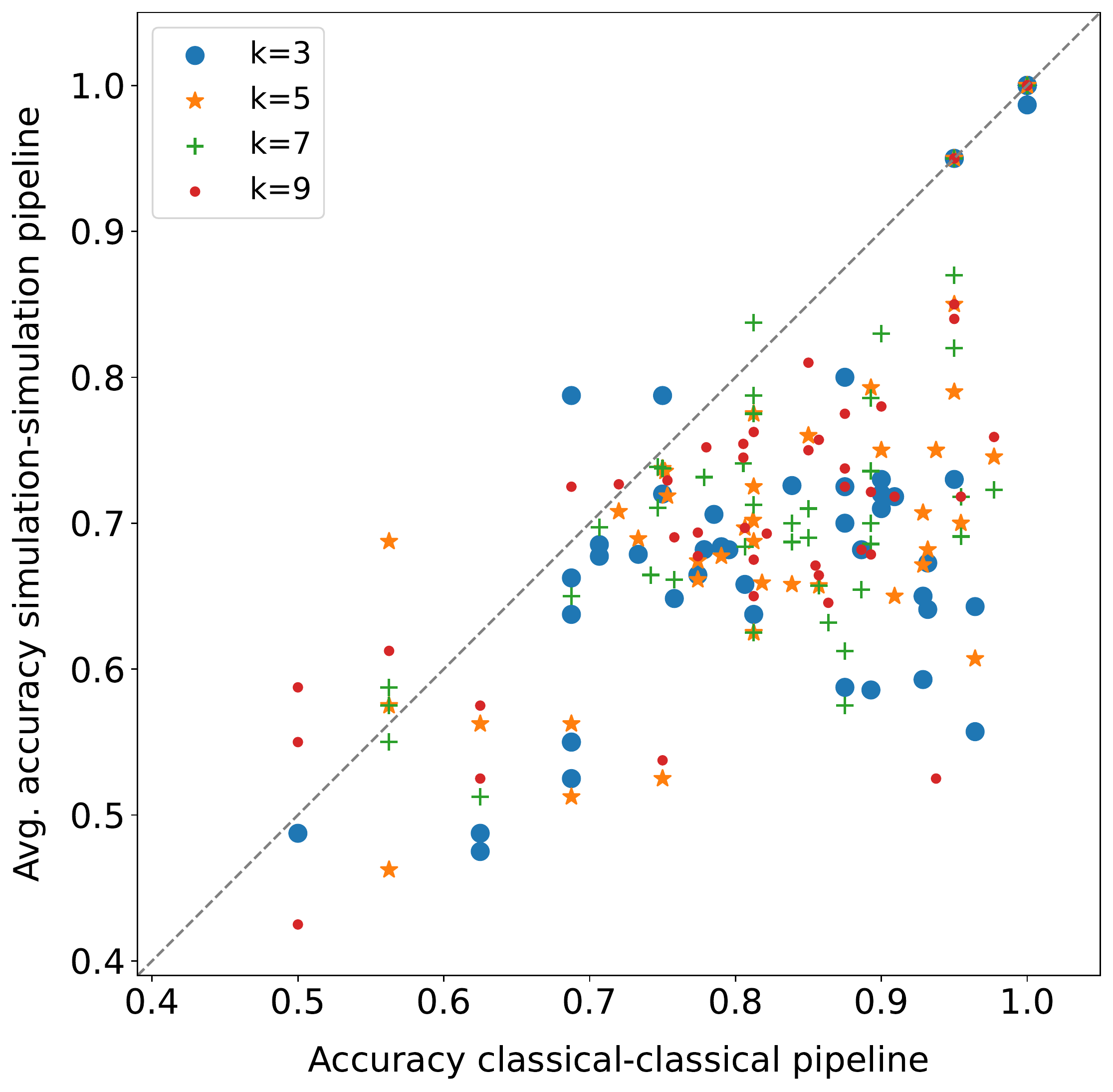}
      \caption{}
      \label{fig:cl-cl-vs-si-si}
    \end{subfigure}
    \caption{Execution modality comparison on \textit{15 qubits} datasets for the quantum pipeline. Each point represents the accuracy obtained in a fold (or its average across runs).}
    \label{fig:pipeline-comp}
        
    \vspace{10pt}
    
    \hypersetup{hidelinks}
    \begin{tabular}{c|c|c|c|c}
		 & \textbf{k=3} & \textbf{k=5} & \textbf{k=7} & \textbf{k=9} \\ \hline
		\cref{fig:cl-cl-vs-sv-sv} & 1.000 & 0.276 & 1.000 & 0.655 \\ \hline
		\cref{fig:cl-cl-vs-si-cl} & 8.80E-08 & 1.18E-07 & 2.46E-06 & 4.01E-06 \\ \hline
		\cref{fig:cl-cl-vs-cl-si} & 0.414 & 0.052 & 0.486 & 0.092 \\ \hline
		\cref{fig:cl-cl-vs-si-si} & 1.04E-07 & 2.03E-07 & 2.27E-07 & 4.60E-07 \\
	\end{tabular}
    \captionof{table}{Wilcoxon signed-rank test ($\alpha\,{=}\,0.05$) applied to the fold accuracy distributions shown in \cref{fig:pipeline-comp}. The values reported in the table are the p-values obtained.}
    \label{tab:pipeline-comp-wilcox}
\end{figure}

\subsubsection{Execution Modalities Comparison (Quantum Pipeline)}
\label{subsubsec:exec-mod-comp-q-pipeline}
Some comparisons between execution modalities for the quantum pipeline on the \textit{15 qubits} datasets are shown in \cref{fig:pipeline-comp}. In these plots, each point represents the accuracy obtained in a fold (or its average across runs). In particular, the \textit{statevector - statevector} modality turns out to be equivalent to the \textit{classical - classical} (\cref{fig:cl-cl-vs-sv-sv}), as expected. Indeed, it is an ideal execution (with an infinite number of runs and without noise) of the quantum circuits implementing the quantum $k$-NN and the quantum binary classifier, thus, it should be equivalent in accuracy to its classical counterpart. The few deviations (from the main diagonal) in \cref{fig:cl-cl-vs-sv-sv} are due to two aspects: the different policies used by the two modalities to select the nearest neighbors in the case of a distance tie; the presence of instances with identical features and different class labels in the \textit{02\_transfusion} dataset. Despite that, the Wilcoxon signed-rank test has confirmed that the difference between the two modalities is not statistically significant (\cref{tab:pipeline-comp-wilcox}). It is also worth recalling that the advantage of the considered quantum models/pipelines with respect to their classical counterparts lies in the execution time and not in the accuracy. The scatter plots comparing the \textit{classical - classical} modality with the \textit{statevector - classical} and the \textit{classical - statevector} (not reported) are identical or almost identical (absence of deviant points) to \cref{fig:cl-cl-vs-sv-sv}.

Although the quantum pipeline is equivalent (in accuracy) to the classical one in the ideal case, this is not true when simulated (even more so when executed on a quantum device, which is also noisy). By looking at \cref{fig:cl-cl-vs-si-cl} (and \ref{fig:cl-cl-vs-si-si}), it turns out that simulating the quantum $k$-NN with 1024 shots (i.e., measurements) adversely affects the performance of the pipeline, independently from the $k$ value. Indeed, almost all points lie below the main diagonal, and the difference is statistically significant for all $k$ values, as reported in \cref{tab:pipeline-comp-wilcox}. In practice, the implemented quantum $k$-NN is really sensitive to fluctuations in the estimated probabilities, and a higher number of shots is needed to obtain better results. Instead, it seems that simulating only the quantum classifier with 1024 shots (\cref{fig:cl-cl-vs-cl-si}) does not have a significant impact on the pipeline performance (the difference is not statistically significant). Nevertheless, the effective usage of the model following the classical (and, thus, the \textit{statevector}) $k$-NN is very low in the datasets used. Most times, all the elements selected by the classical $k$-NN belong to the same class, and therefore it is not possible to draw definitive conclusions about the subsequent classifier. The average \textit{usage on dataset} of the second model for the cosine distance metric and each $k$ value is reported in \cref{tab:avg-2nd-model-usage-cos}.

\begin{table}[b]
    \centering
    \begin{subtable}[t]{.495\textwidth}
        \centering
        \begin{tabular}{c|c|c}
            \textbf{k} & \textbf{15 qubits} & \textbf{32 qubits} \\ \hline
            3 & 0.245 $\pm$ 0.204 & 0.375 $\pm$ 0.128 \\ \hline
            5 & 0.357 $\pm$ 0.290 & 0.590 $\pm$ 0.160 \\ \hline
            7 & 0.406 $\pm$ 0.323 & 0.690 $\pm$ 0.158 \\ \hline
            9 & 0.461 $\pm$ 0.356 & 0.758 $\pm$ 0.137 \\
        \end{tabular}
        \caption{Cosine.} 
        \label{tab:avg-2nd-model-usage-cos}
    \end{subtable}
    \begin{subtable}[t]{.495\textwidth}
        \centering
        \begin{tabular}{c|c|c}
            \textbf{k} & \textbf{15 qubits} & \textbf{32 qubits} \\ \hline
            3 & 0.218 $\pm$ 0.229 & 0.402 $\pm$ 0.133 \\ \hline
            5 & 0.298 $\pm$ 0.305 & 0.617 $\pm$ 0.187 \\ \hline
            7 & 0.346 $\pm$ 0.346 & 0.686 $\pm$ 0.188 \\ \hline
            9 & 0.381 $\pm$ 0.371 & 0.755 $\pm$ 0.166 \\
        \end{tabular}
        \caption{Euclidean.} 
        \label{tab:avg-2nd-model-usage-euc}
    \end{subtable}
    \caption{Average \textit{usage on dataset} of the second model for the pipelines including the classical (or statevector) $k$-NN with cosine distance (a) and Euclidean distance (b). The \textit{usage on dataset} is 1 when the second model is always employed.}
    \label{tab:avg-2nd-model-usage}
\end{table}

\begin{figure}[t!]
    \centering
    \includegraphics[width=0.45\textwidth]{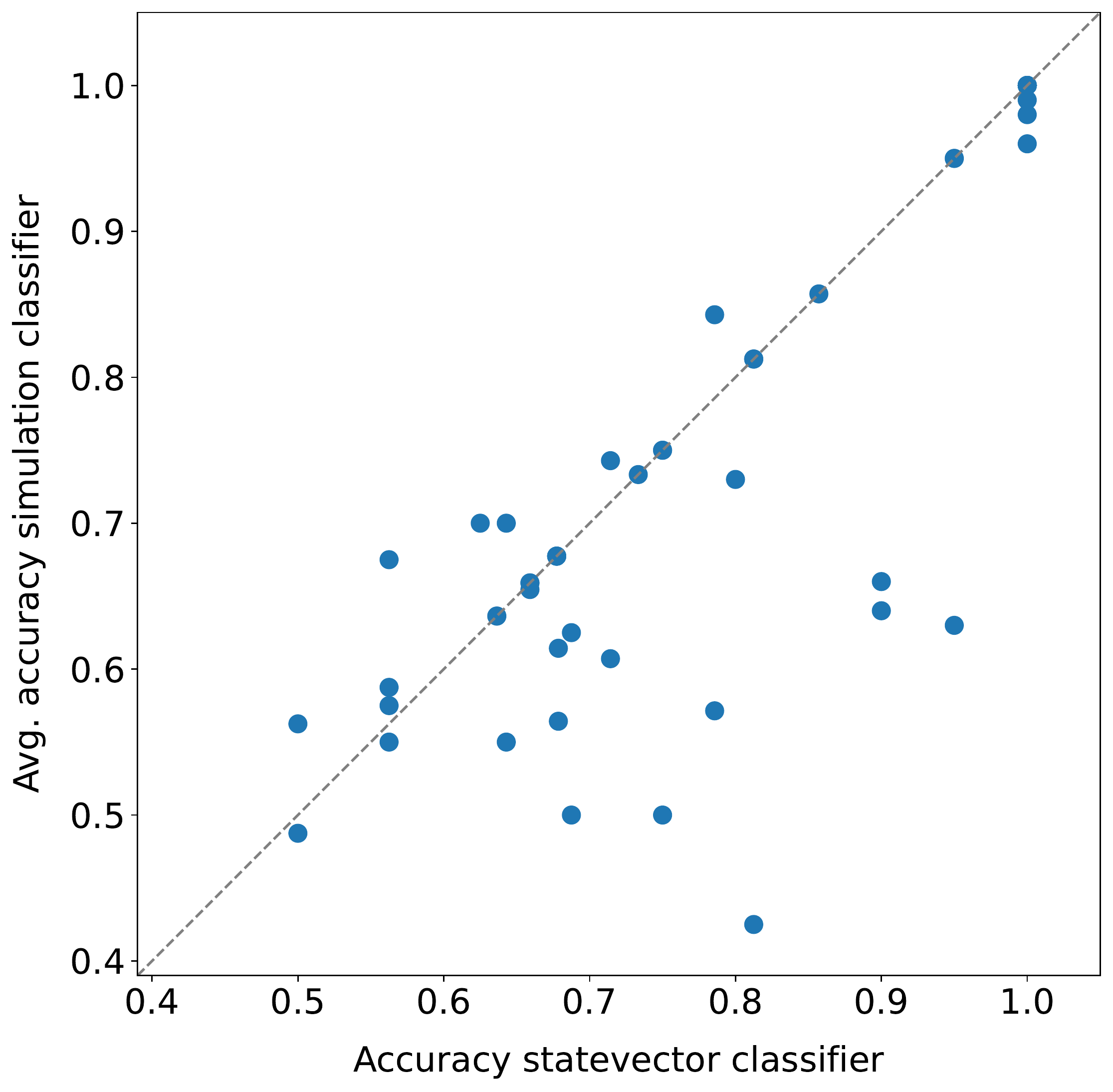}
    \caption{Execution modality comparison on \textit{15 qubits} datasets (the \textit{02\_transfusion} dataset is not present) for the quantum binary classifier. Each point represents the accuracy obtained in a fold (or its average across runs). The p-value obtained by applying the Wilcoxon signed-rank test ($\alpha\,{=}\,0.05$) to the fold accuracy distributions is $0.016$.}
    \label{fig:qbc-comp}
\end{figure}

\subsubsection{Execution Modalities Comparison (Quantum Binary Classifier Alone)}
\label{subsubsec:exec-mod-comp-qbc}
The comparison between \textit{statevector} and \textit{simulation} modalities for the quantum binary classifier alone on the \textit{15 qubits} datasets is displayed in \cref{fig:qbc-comp}. The \textit{classical} modality has not been taken into account due to the effective equivalence with respect to \textit{statevector}; in addition, the \textit{02\_transfusion} dataset has not been included since an additional qubit would have been required. Also in this case, each dot represents the accuracy obtained in a fold (or its average across runs). Here, it is possible to observe that the quantum binary classifier as well is affected by the probability fluctuations and, more practically, by the number of shots. Indeed, the \textit{simulation} performance is worse than the \textit{statevector}'s overall, and the difference is confirmed by the Wilcoxon signed-rank test (p-value$=0.016$). 

\begin{figure}[t!]
    \centering
    \begin{subfigure}{0.45\textwidth}
      \centering
      \includegraphics[width=\textwidth]{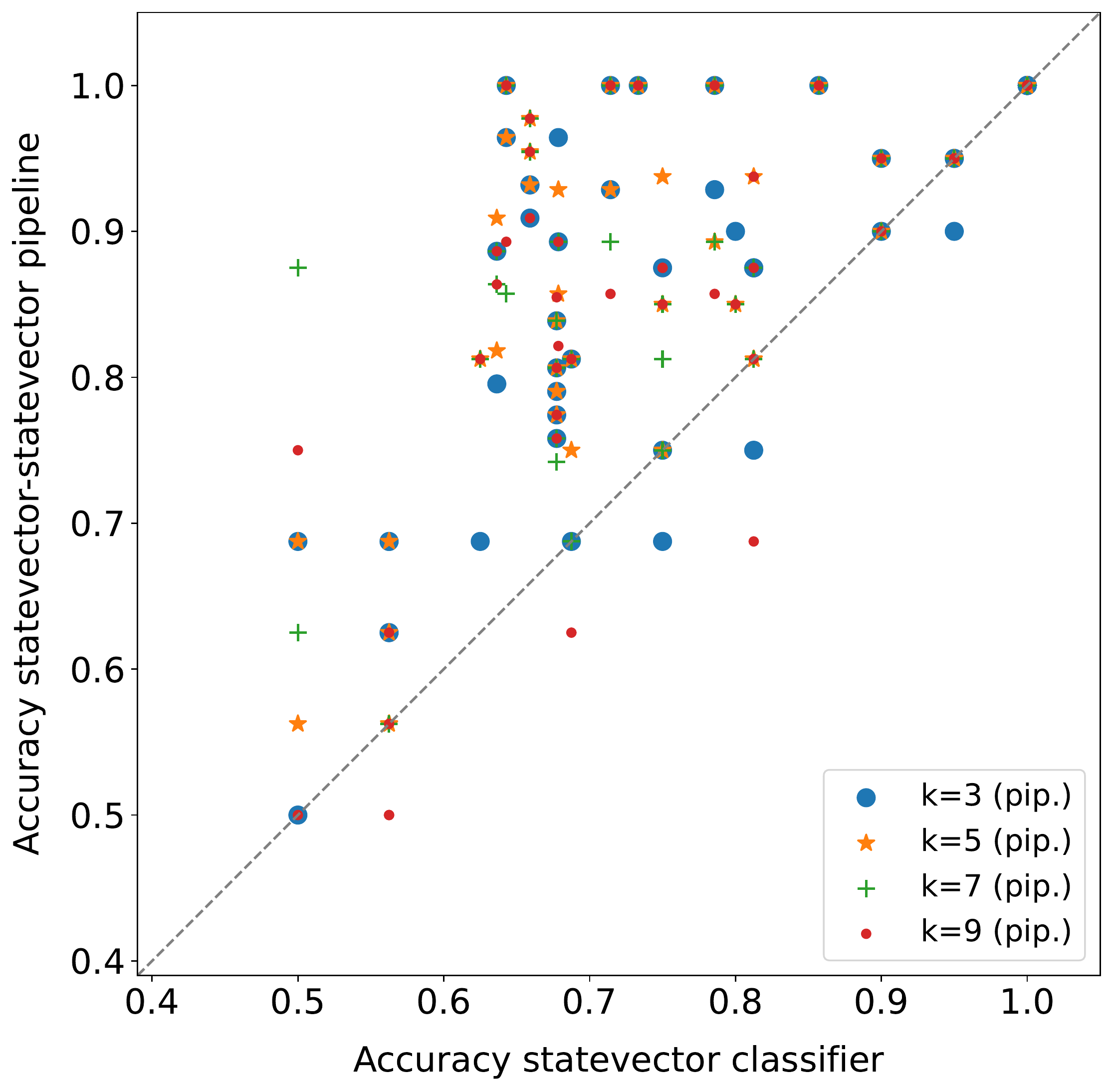}
      \caption{}
      \label{fig:sv-vs-sv-sv}
    \end{subfigure}
    \begin{subfigure}{0.45\textwidth}
      \centering
      \includegraphics[width=\textwidth]{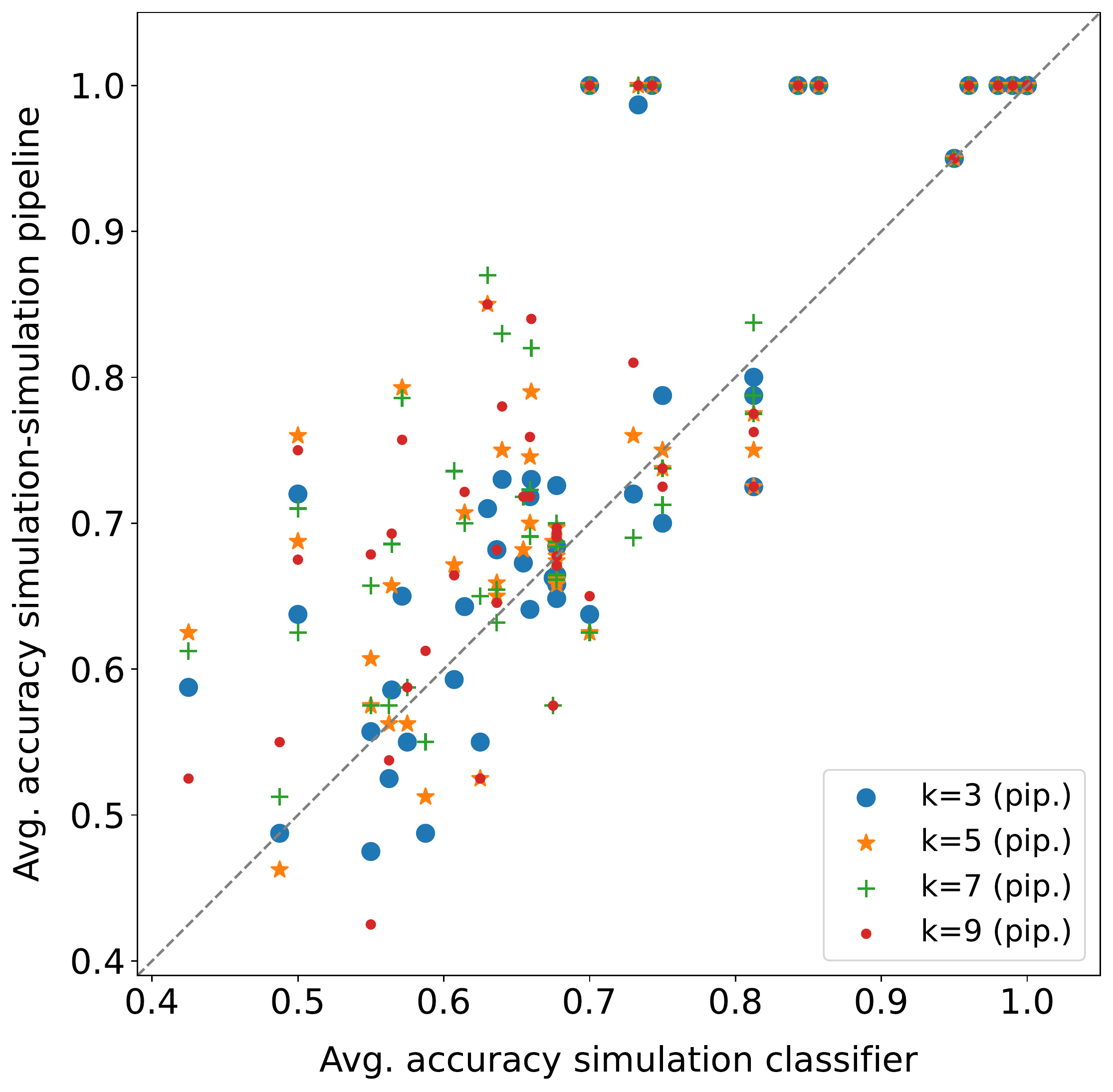}
      \caption{}
      \label{fig:si-vs-si-si}
    \end{subfigure}
    \caption{Quantum pipeline - quantum binary classifier comparison on common \textit{15 qubits} datasets, with each point representing the accuracy obtained in a fold (or its average across runs). The $k$ values refer only to the pipeline.}
    \label{fig:pipeline-vs-qbc}
    
    \hypersetup{hidelinks}
    \begin{tabular}{c|c|c|c|c}
		 & \textbf{k=3} & \textbf{k=5} & \textbf{k=7} & \textbf{k=9} \\ \hline
		\cref{fig:sv-vs-sv-sv} & 5.39E-07 & 7.88E-07 & 2.55E-06 & 2.98E-06 \\ \hline
		\cref{fig:si-vs-si-si} & 0.042 & 0.001 & 3.35E-04 & 5.84E-04 \\
	\end{tabular}
    \captionof{table}{Wilcoxon signed-rank test ($\alpha\,{=}\,0.05$) applied to the fold accuracy distributions shown in \cref{fig:pipeline-vs-qbc}. The values reported in the table are the p-values obtained.}
    \label{tab:pipeline-vs-qbc-wilcox}
\end{figure}

\subsubsection{Quantum Pipeline - Quantum Binary Classifier Comparison}
\label{subsubsec:q-pipeline-qbc-comp}
The comparison between the quantum pipeline and the quantum binary classifier on the \textit{15 qubits} datasets is illustrated in \cref{fig:pipeline-vs-qbc}; the structure of the scatter plots is the same as that of the charts in \cref{fig:pipeline-comp}, although here the \textit{02\_transfusion} dataset is not present to allow the comparison and the $k$ values in the legend refer only to the pipeline. As shown in \cref{fig:sv-vs-sv-sv}, the \textit{statevector - statevector} pipeline performs better than the \textit{statevector} classifier for all $k$ values. In detail, it statistically outperforms the classifier alone (\cref{tab:pipeline-vs-qbc-wilcox}). Actually, there are some folds in which the classifier alone achieves the highest accuracy, but they represent a clear minority. This confirms the effectiveness of applying a quantum locality technique such as the quantum $k$-NN as a preliminary step of a quantum classifier, and more in general, the worth of locality. Instead, the pipeline superiority is far less marked when the simulated versions of the pipeline and the quantum binary classifier are considered (\cref{fig:si-vs-si-si}). Indeed, it turns out that the quantum pipeline, mainly due to the implemented quantum $k$-NN, is far more negatively affected by probability fluctuations than the classifier alone. Nevertheless, it still manages to perform better overall, independently from the number of nearest neighbors selected (the difference is statistically significant also in this case). Eventually, looking at the different $k$ values in these two plots, there is no dominant one; indeed, the best $k$ value is typically dataset-dependent.

\subsubsection{Dataset Sizes Comparison}
\label{subsubsec:dataset-size-comparison}
The effect of the dataset size on the performance of the quantum pipeline and the quantum binary classifier is analysed in \cref{fig:dataset-size-comp}. In detail, only the three datasets in \cref{tab:datasets} having both \textit{15 qubits} and \textit{32 qubits} size have been considered for this plot. Moreover, since a fold by fold comparison would not make sense in this case, each point in the chart represents the mean fold accuracy on a dataset (or its average across runs). Finally, the results for all $k$ values have been included for the pipelines; as a consequence, the number of pipeline points in the plot is four times higher with respect to the classifier alone. In practice, a larger dataset tends to have a beneficial effect on the performance of the pipeline in the ideal case (\textit{statevector - statevector}) and an overall neutral effect on its simulation: the occurrences of improvement and worsening are the same in the latter. Instead, overall, the performance of the quantum classifier alone worsens in both cases. This represents another point in favor of the quantum pipeline, which turns out to be able to take advantage of a larger number of samples. Actually, the difference is statistically significant in the case of the \textit{statevector - statevector} pipeline, whereas it is not in the others, as shown in Table~\ref{tab:dataset-size-comp-wilcox}. However, in this case, the low number of points must also be taken into account.

\begin{figure}[t!]
    \centering
    \begin{subfigure}{0.45\textwidth}
      \centering
        \includegraphics[width=\textwidth]{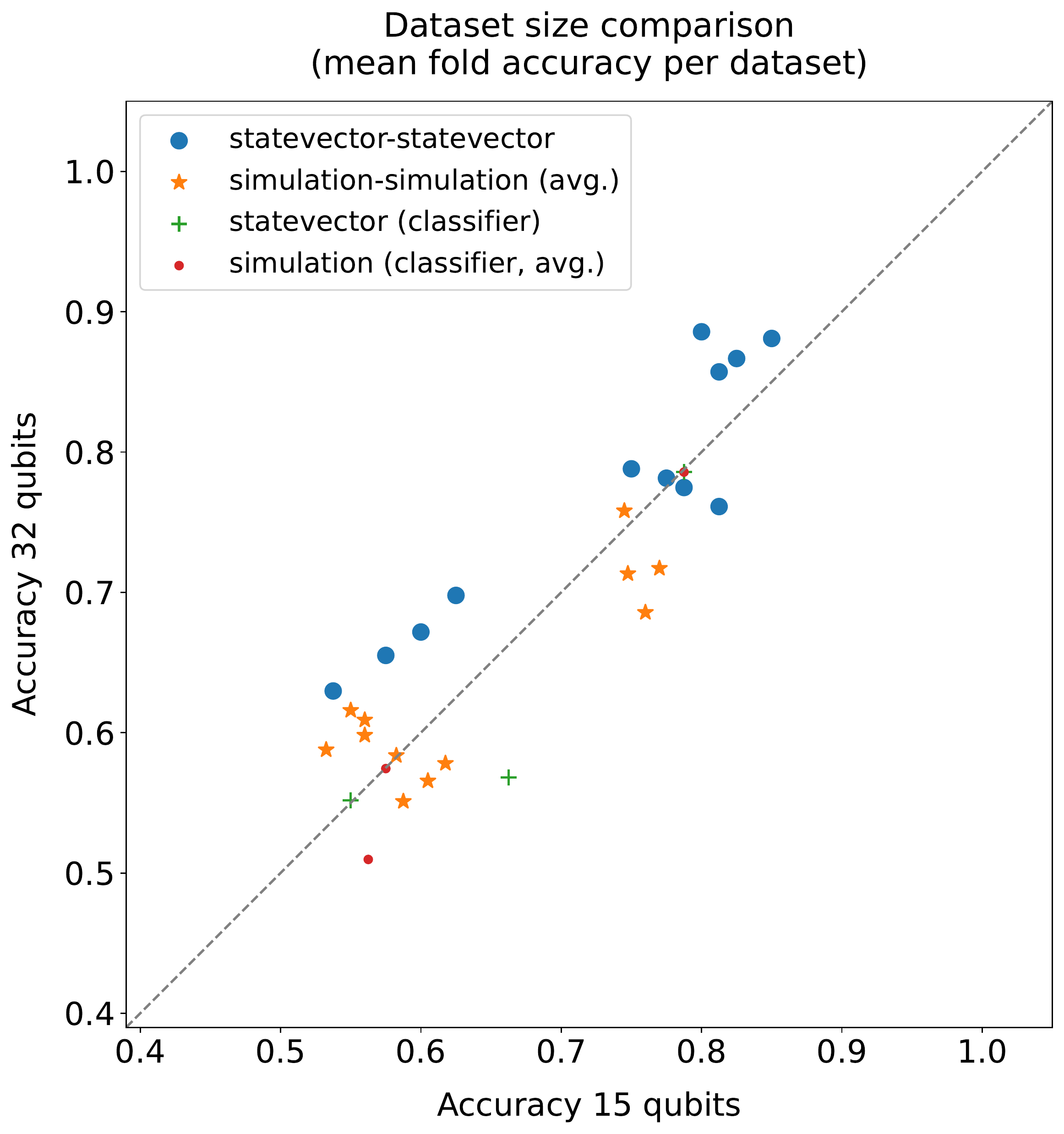}
        \caption{}
        \label{fig:dataset-size-comp}
    \end{subfigure}
    \begin{subfigure}{0.45\textwidth}
      \centering
      \includegraphics[width=\textwidth]{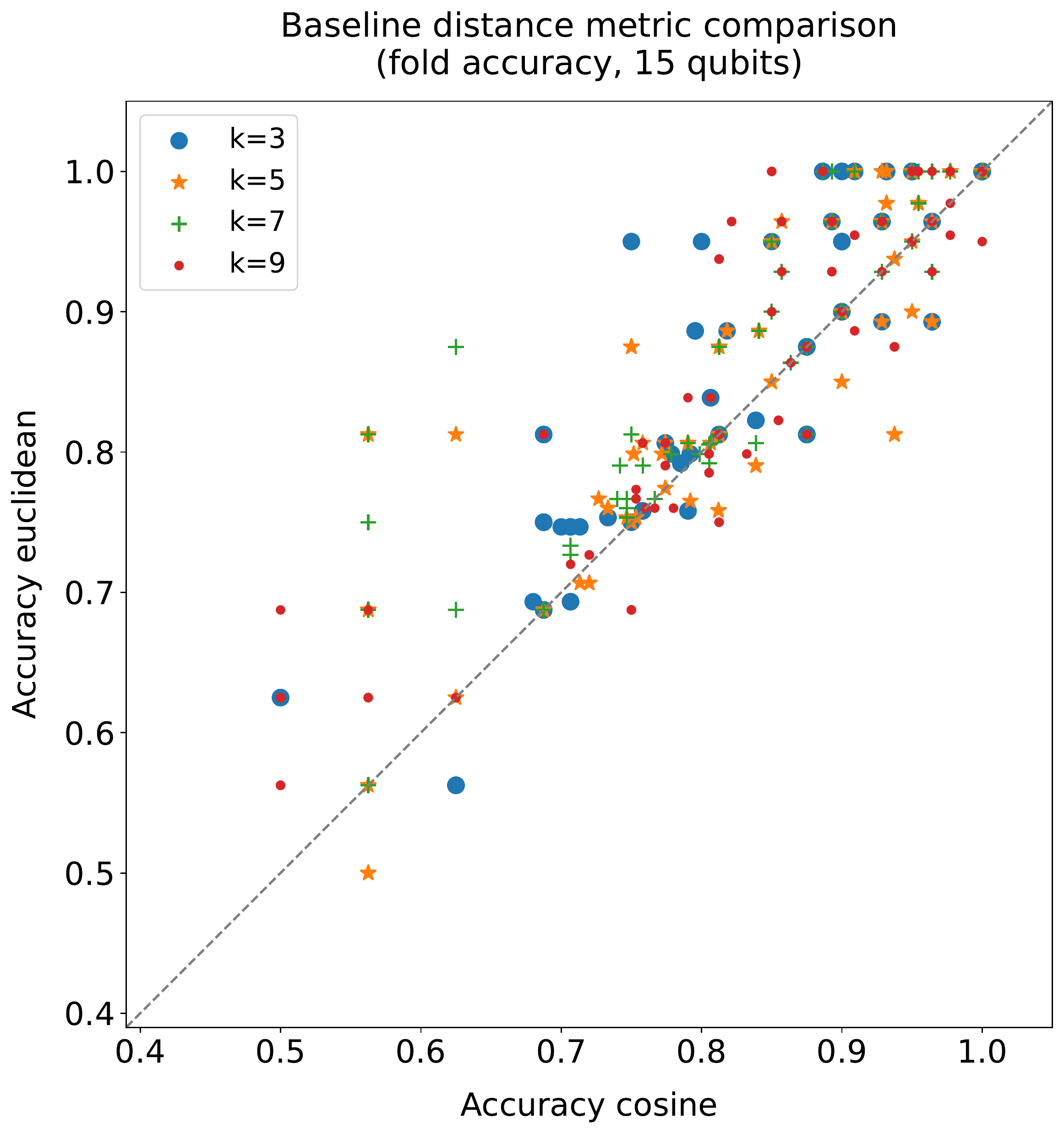}
      \caption{}
      \label{fig:dist-metric-comp}
    \end{subfigure}
    \caption{Dataset size comparison (a), with each point representing the mean fold accuracy obtained on a dataset (or its average across runs); the pipeline comparisons include all $k$ values. Distance metric comparison in $k$-NN-based baseline methods (\textit{k-NN}, \textit{k-NN + classifier}, \textit{k-NN + SVM Gaussian}, \textit{k-NN + SVM linear}) on \textit{15 qubits} datasets (b), with each point representing the accuracy obtained in a fold.}
    \label{fig:dataset-size-n-dist-metric-comp}
    
    \begin{subtable}[]{.40\textwidth}
        \centering
        \begin{tabular}{c|c}
             & \textbf{p-value} \\ \hline
    		\textbf{\textit{statev.-statev.}} & 0.016 \\ \hline
    		\textbf{\textit{sim.-sim.}} & 0.910 \\ \hline
    		\textbf{\textit{statev.}} & 0.750 \\ \hline
    		\textbf{\textit{sim.}} & 0.250 \\
    	\end{tabular}
    	\caption{Dataset size.}
        \label{tab:dataset-size-comp-wilcox}
    \end{subtable}
    \begin{subtable}[]{.40\textwidth}
        \centering
        \begin{tabular}{c|c}
             & \textbf{p-value} \\ \hline
    		\textbf{k=3} & 9.85E-10 \\ \hline
    		\textbf{k=5} & 8.33E-08 \\ \hline
    		\textbf{k=7} & 7.65E-15 \\ \hline
    		\textbf{k=9} & 4.86E-08 \\
    	\end{tabular}
    	\caption{Distance metric.}
        \label{tab:dist-metric-comp-wilcox}
    \end{subtable}
    \captionof{table}{Wilcoxon signed-rank test ($\alpha\,{=}\,0.05$) applied to the mean fold accuracy distributions shown in \cref{fig:dataset-size-comp} (a). Same test applied to the fold accuracy distributions shown in \cref{fig:dist-metric-comp} (b).}
    \label{tab:dataset-size-n-dist-metric-comp-wilcox}
    \addtocounter{figure}{-1}
\end{figure}

\subsubsection{Distance Metrics Comparison}
\label{subsubsec:dist-comp}
As mentioned in \cref{subsec:methods}, two distance metrics, namely, the cosine and the Euclidean distances, have been evaluated for the baseline methods based on the $k$-NN algorithm, i.e., the $k$-NN, the \textit{k-NN + classifier}, and the \textit{k-NN + SVM} with both Gaussian and linear kernels. The comparison between these two metrics on the \textit{15 qubits} datasets is shown in \cref{fig:dist-metric-comp}, with each point representing the accuracy obtained in a fold by one of the four just cited methods. Basically, the Euclidean distance statistically outperforms (Table~\ref{tab:dist-metric-comp-wilcox}) the cosine one on the datasets used for all $k$ values. Therefore, it would be advantageous to have a quantum $k$-NN version based on that metric.

\begin{figure}[b!]
    \centering
    \begin{subfigure}{0.45\textwidth}
      \centering
      \includegraphics[width=\textwidth]{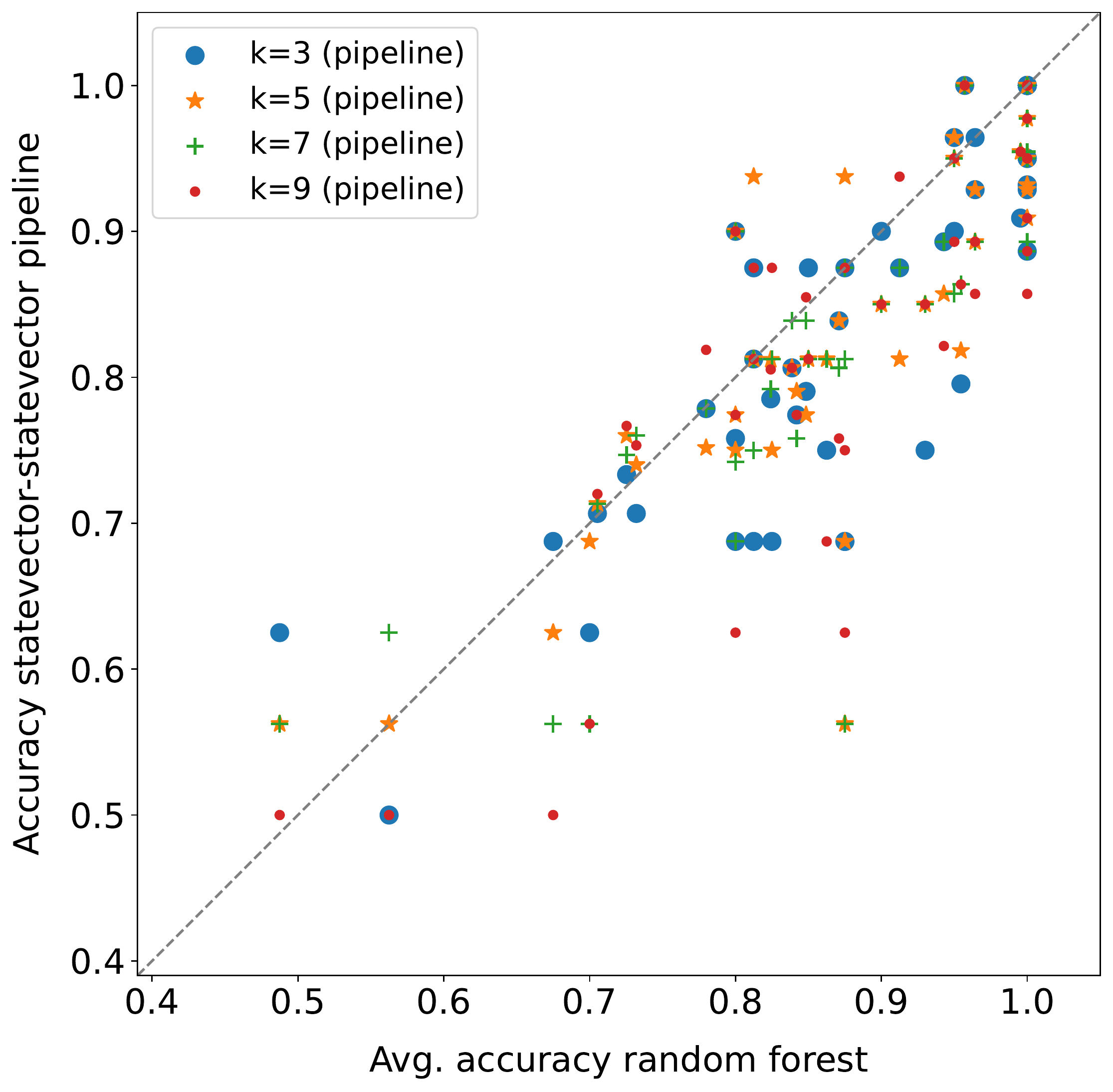}
      \caption{}
      \label{fig:random-forest-vs-sv-sv}
    \end{subfigure}
    \begin{subfigure}{0.45\textwidth}
      \centering
      \includegraphics[width=\textwidth]{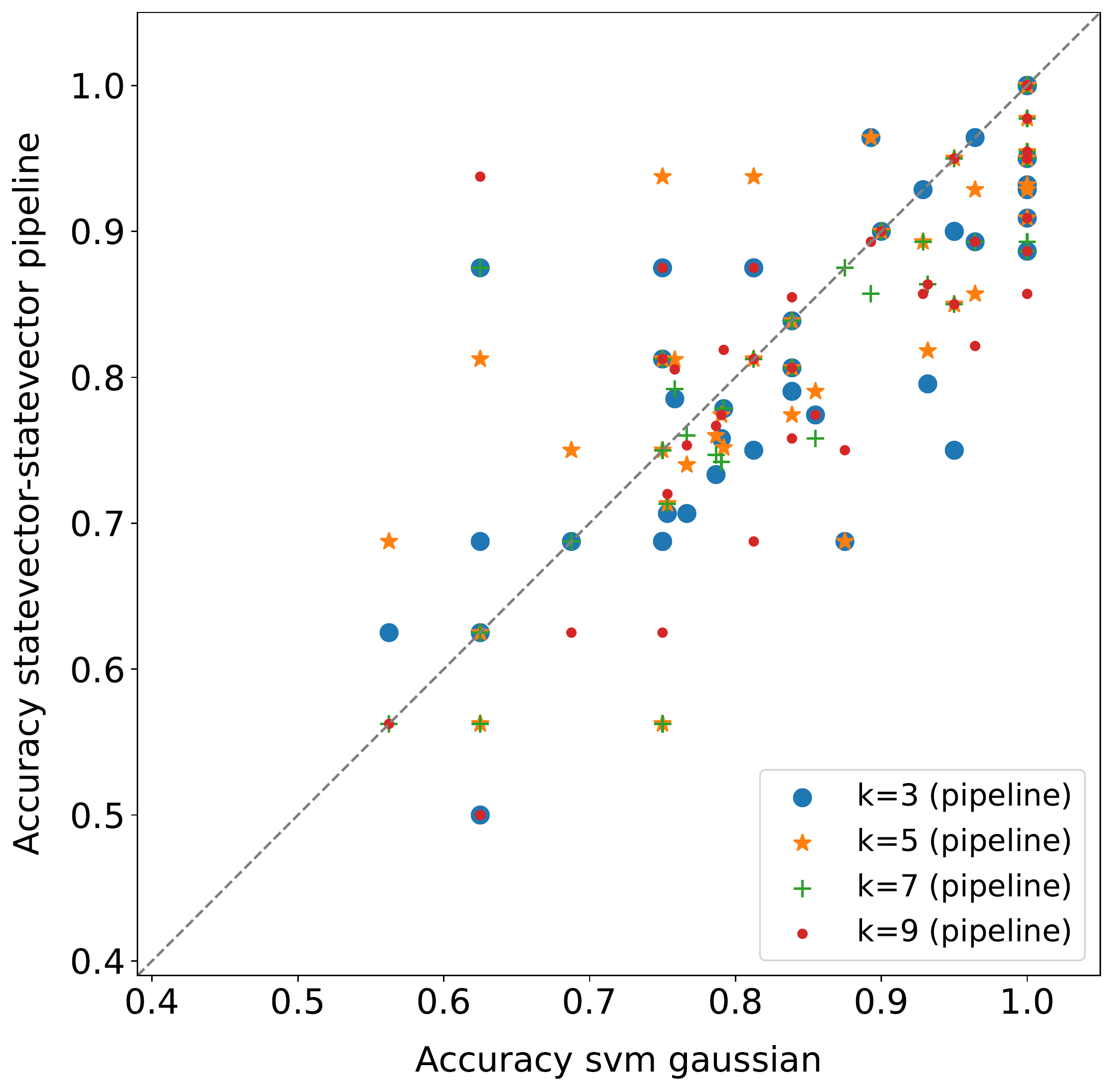}
      \caption{}
      \label{fig:svm-rbf-vs-sv-sv}
    \end{subfigure}
    \caption{Quantum pipeline - baseline methods comparison on \textit{15 qubits} datasets (the pipeline modality is \textit{statevector - statevector}). Each point in these plots represents the accuracy obtained in a fold (or its average across runs); the $k$-values refer only to the pipeline.}
    \label{fig:baseline-vs-pipeline}
    
    \hypersetup{hidelinks}
    \begin{tabular}{c|c|c|c|c}
		 & \textbf{k=3} & \textbf{k=5} & \textbf{k=7} & \textbf{k=9} \\ \hline
		\cref{fig:random-forest-vs-sv-sv} & 1.54E-04 & 0.001 & 1.24E-04 & 1.88E-04 \\ \hline
		\cref{fig:svm-rbf-vs-sv-sv} & 0.020 & 0.106 & 0.003 & 0.004 \\
	\end{tabular}
    \captionof{table}{Wilcoxon signed-rank test ($\alpha\,{=}\,0.05$) applied to the fold accuracy distributions shown in \cref{fig:baseline-vs-pipeline}. The values reported in the table are the p-values obtained.}
    \label{tab:baseline-vs-pipeline-wilcox}
\end{figure}

\begin{figure}[p]
    \centering
    \begin{subfigure}{0.45\textwidth}
      \centering
      \includegraphics[width=\textwidth]{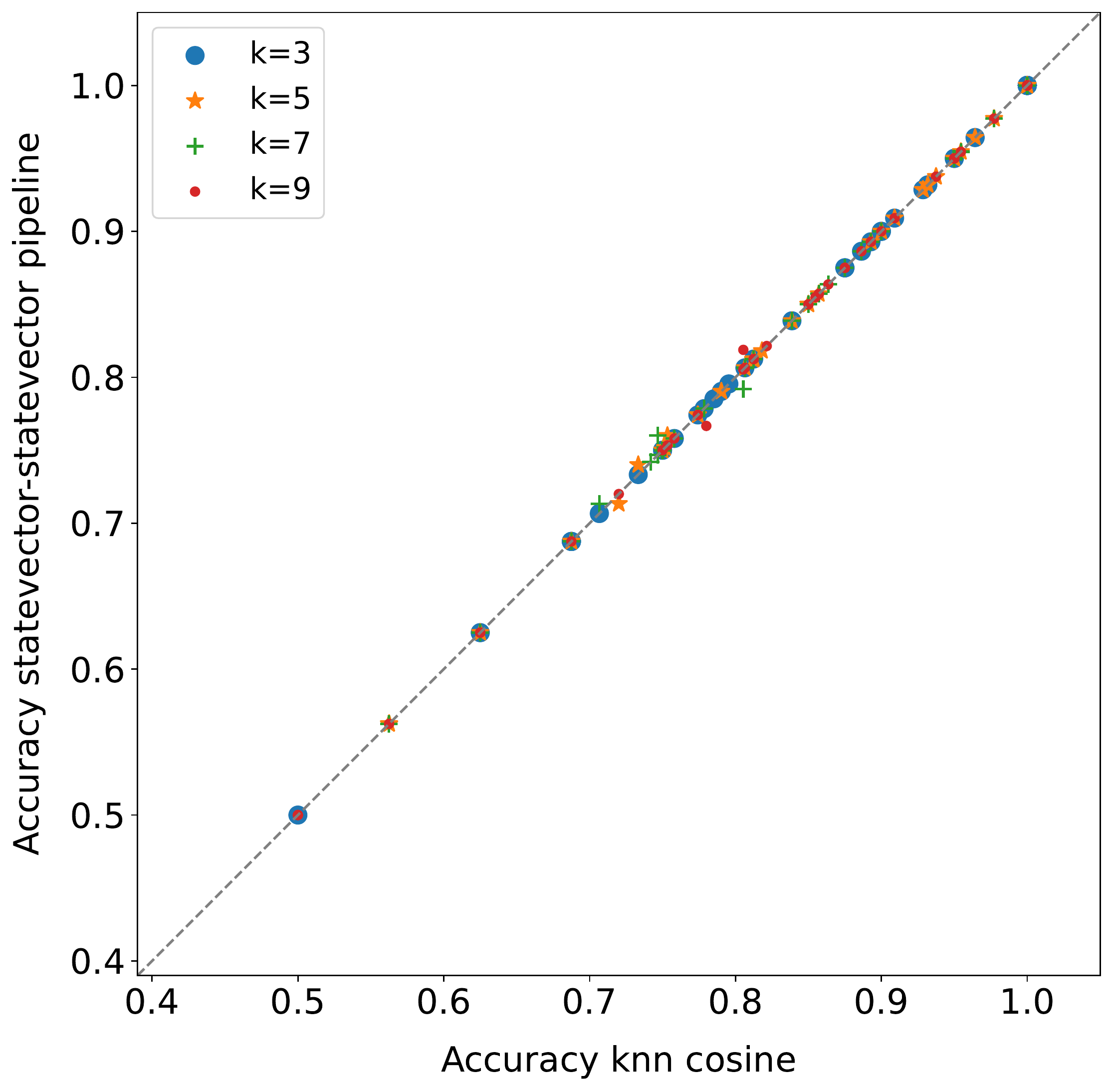}
      \caption{}
      \label{fig:knn-cos-vs-sv-sv}
    \end{subfigure}
    \begin{subfigure}{0.45\textwidth}
      \centering
      \includegraphics[width=\textwidth]{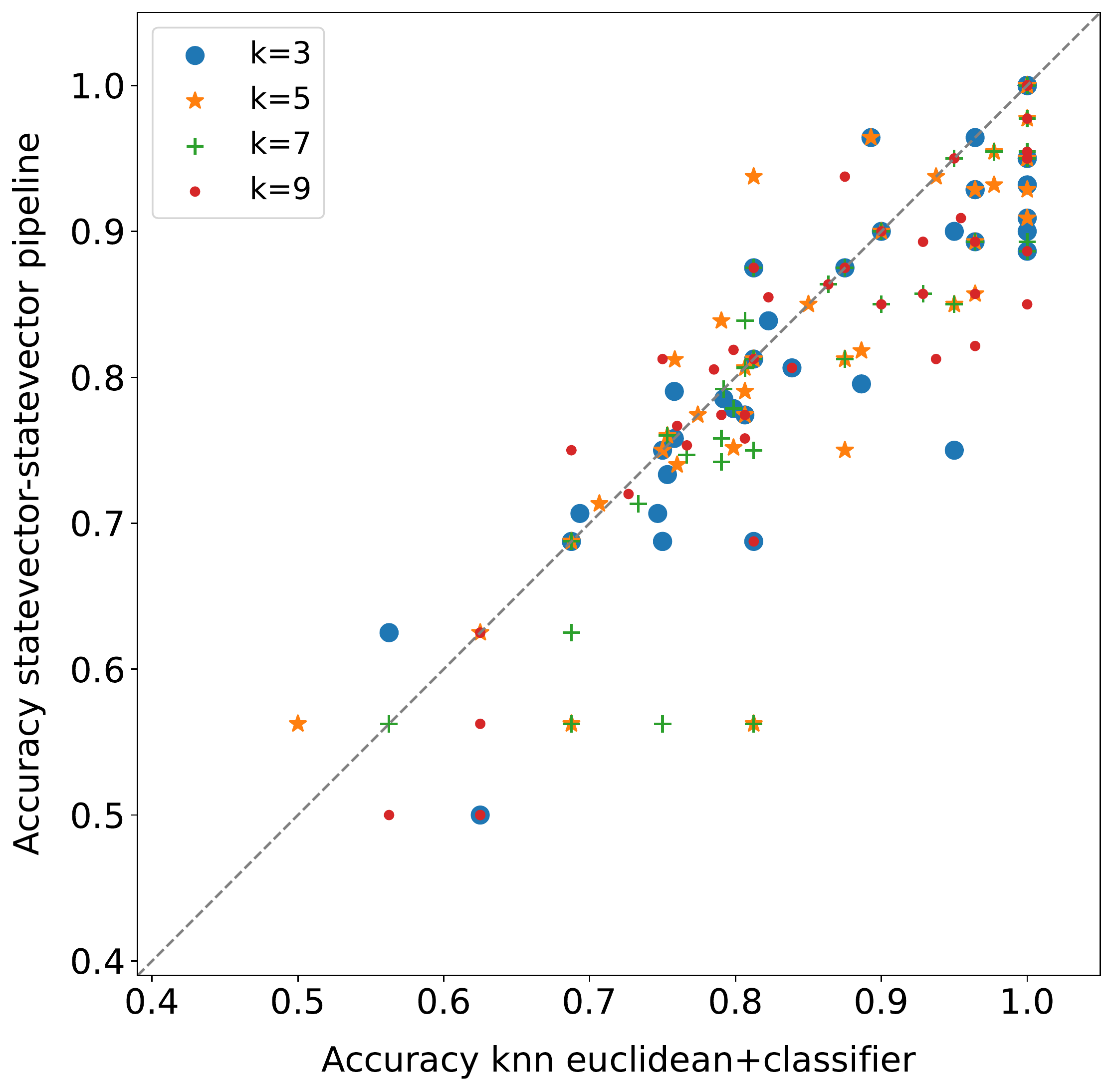}
      \caption{}
      \label{fig:knn-euc+classif-vs-sv-sv}
    \end{subfigure}
    \begin{subfigure}{0.45\textwidth}
      \centering
      \includegraphics[width=\textwidth]{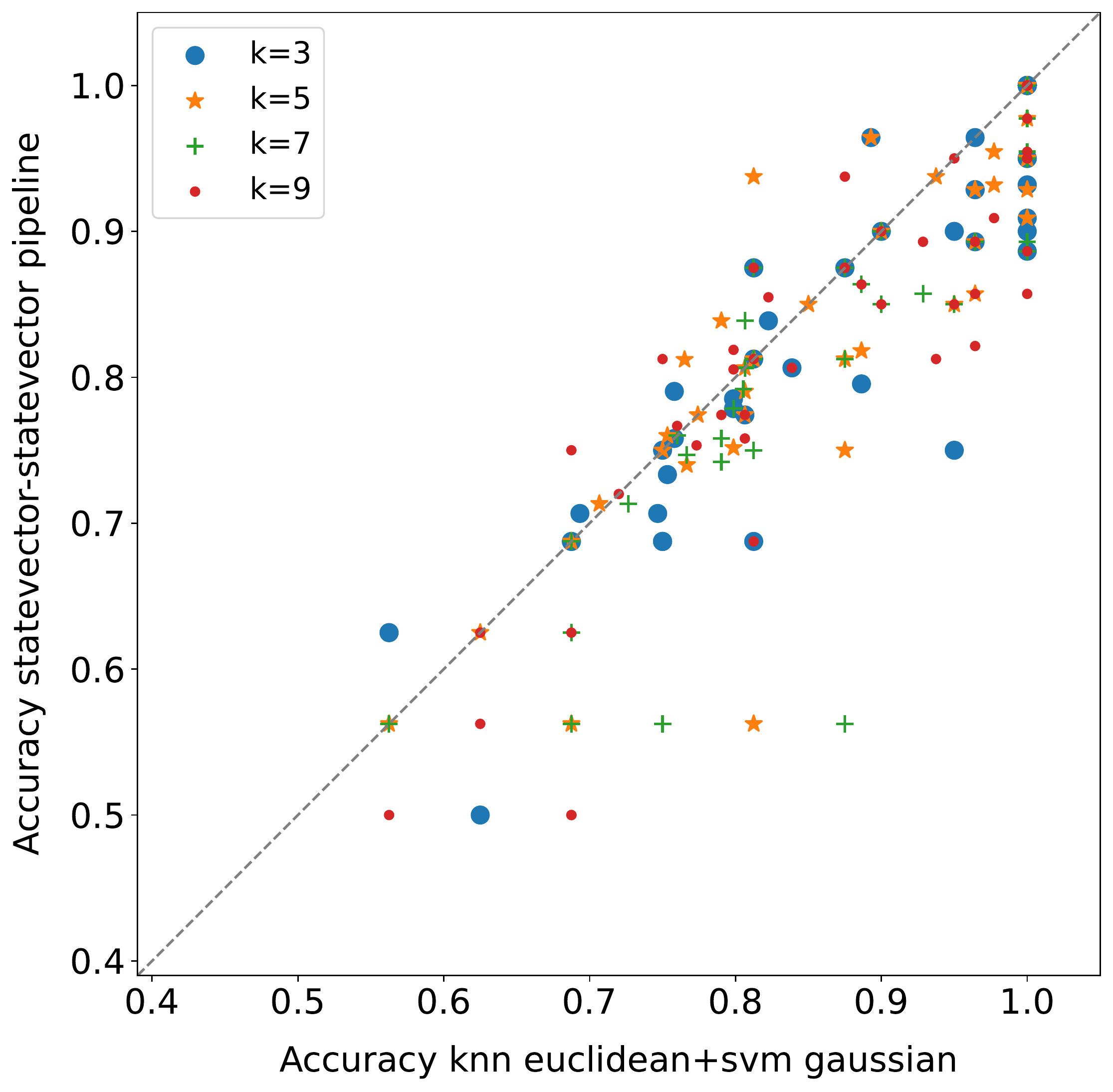}
      \caption{}
      \label{fig:knn-euc+svm-rbf-vs-sv-sv}
    \end{subfigure}
    \caption{Quantum pipeline - ($k$-NN-based) baseline methods comparison on \textit{15 qubits} datasets. Each point in these plots represents the accuracy obtained in a fold.}
    \label{fig:knn-baseline-vs-pipeline}
    
    \hypersetup{hidelinks}
    \begin{tabular}{c|c|c|c|c}
		 & \textbf{k=3} & \textbf{k=5} & \textbf{k=7} & \textbf{k=9} \\ \hline
		\cref{fig:knn-cos-vs-sv-sv} & 1.000 & 0.276 & 1.000 & 0.655 \\ \hline
		\cref{fig:knn-euc+classif-vs-sv-sv} & 0.003 & 0.006 & 5.22E-05 & 0.001 \\ \hline
		\cref{fig:knn-euc+svm-rbf-vs-sv-sv} & 0.003 & 0.002 & 2.15E-05 & 4.36E-04 \\
	\end{tabular}
    \captionof{table}{Wilcoxon signed-rank test ($\alpha\,{=}\,0.05$) applied to the fold accuracy distributions shown in \cref{fig:knn-baseline-vs-pipeline}. The values reported in the table are the p-values obtained.}
    \label{tab:knn-baseline-vs-pipeline-wilcox}
\end{figure}

\subsubsection{Quantum Pipeline - Baseline Methods Comparison}
\label{subsubsec:q-pipeline-baseline-comp}
Some comparisons between the \textit{statevector - statevector} pipeline and baseline methods on the \textit{15 qubits} datasets are displayed in \cref{fig:baseline-vs-pipeline,fig:knn-baseline-vs-pipeline}. As usual, each point represents the accuracy obtained in a fold (or its average across runs); moreover, the $k$ values in the legends of Figures \ref{fig:random-forest-vs-sv-sv} and \ref{fig:svm-rbf-vs-sv-sv} refer only to the pipeline. In practice, the random forest achieves better results than the quantum pipeline for all $k$-values (\cref{fig:random-forest-vs-sv-sv}), and the same applies to the best SVM, i.e., the SVM with the Gaussian kernel (\cref{fig:svm-rbf-vs-sv-sv}). The difference turns out to be statistically significant in both cases, as shown in \cref{tab:baseline-vs-pipeline-wilcox}, with the exception of $k=5$ for the SVM - pipeline comparison. Instead, the SVM with the linear kernel (not shown here) performs just slightly better than the pipeline, and the difference is not statistically significant.

Regarding the $k$-NN, the version with cosine distance metric turns out to be equivalent to the \textit{statevector - statevector} pipeline (\cref{fig:knn-cos-vs-sv-sv}, the few deviations are the same as the ones in \cref{fig:cl-cl-vs-sv-sv}); as shown in \cref{tab:knn-baseline-vs-pipeline-wilcox}, the difference is not statistically significant. Therefore, it is also equivalent to the \textit{classical - classical} pipeline (see \cref{subsubsec:exec-mod-comp-q-pipeline}). This means that, on the considered datasets, a label assignment based on the $k$ nearest neighbors extracted using the cosine distance produces the same outcome by applying either a majority voting or the binary classifier based on the cosine similarity. Indeed, the lack of unit-norm normalization of the input data in the baseline methods does not affect the cosine distance (or the cosine similarity), which intrinsically normalizes them. However, it is also worth remarking on the low usage of the models following the nearest neighbors extraction (\cref{tab:avg-2nd-model-usage-cos}). Instead, the $k$-NN with Euclidean distance statistically outperforms the quantum pipeline for all $k$ values, and the resulting scatter plot is identical to \cref{fig:knn-euc+classif-vs-sv-sv} (see \cref{tab:knn-baseline-vs-pipeline-wilcox} for the statistical test results). In fact, the observation made on the label assignment criteria for the cosine distance also holds for the Euclidean distance. Basically, with the same distance metric, the $k$-NN and the \textit{k-NN + classifier} perform equally on the datasets taken into account. As a consequence, the \textit{statevector - statevector} pipeline achieves the same results as the \textit{k-NN + classifier} with cosine distance metric (the corresponding scatter plot is identical to \cref{fig:knn-cos-vs-sv-sv}) and is outperformed by the same model with Euclidean distance for all $k$ values (\cref{fig:knn-euc+classif-vs-sv-sv}). Actually, in the end, the \textit{k-NN + classifier} with cosine distance and the \textit{classical - classical} pipeline represent exactly the same model due to the observation on the unit-norm normalization just made. Finally, concerning the baseline pipelines including the SVM, the best among them is the \textit{k-NN + SVM} with Euclidean distance and Gaussian kernel, which statistically outperforms the \textit{statevector - statevector} pipeline independently from the $k$ value, as shown in \cref{fig:knn-euc+svm-rbf-vs-sv-sv} (the statistical test results are reported in \cref{tab:knn-baseline-vs-pipeline-wilcox}). The other versions of this baseline pipeline still win the comparison, although by a less margin. In particular, the \textit{k-NN + SVM} with cosine distance and linear kernel is almost equivalent to the quantum pipeline, whereas the others are clearly superior. Moreover, the \textit{k-NN + SVM} with Euclidean distance and linear kernel turns out to be equivalent to the \textit{k-NN + classifier} with the same distance metric, which does not hold for the cosine distance (the SVM performs slightly better than the binary classifier in that case). 

Although it is not possible to draw definitive conclusions due to the low effective usage of the model following the $k$ nearest neighbors extraction, the SVM seems to perform better than the cosine similarity classifier in the baseline pipelines (especially with Gaussian kernel); as regards the Euclidean distance metric, the effective usage situation is slightly worse than that of the cosine distance, as reported in \cref{tab:avg-2nd-model-usage-euc}. Hence, it could be beneficial to try to combine a quantum $k$-NN version with a quantum SVM. Eventually, it is worth making a last observation: the classical SVMs tend to outperform the corresponding pipelines (in terms of kernel) with cosine distance and be outperformed by/be equivalent to the corresponding ones with Euclidean distance (but, again, the low usage of the second model in the pipelines must be taken into account).
\vspace{20pt}

\section{Conclusion}
\label{sec:conclusion}
In this paper, we have presented an implementation in Python of a quantum pipeline consisting of a quantum $k$-NN \citep{qknn_implemented} and a quantum binary classifier \citep{binary_classifier_published} and its extensive empirical evaluation. Details about the implementation (based on Qiskit) and the complexity (both theoretical and practical) of the pipeline have been provided, as well as information on the experiments performed, i.e., methods, datasets, and the setup used. First of all, the results have demonstrated the quantum pipeline's equivalence to its classical counterpart in terms of accuracy and its capability of taking advantage of larger datasets (both of them in the ideal case). Moreover, the validity of locality's application to the quantum realm has been confirmed, with the quantum pipeline outperforming the quantum binary classifier in both the ideal and simulated cases. However, the strong sensitivity of the implemented quantum $k$-NN to probability fluctuations has also emerged; indeed, the quantum binary classifier is affected by the same issue to a much lesser extent. In addition, the quantum pipeline has been outperformed by baseline methods like the random forest and the SVMs even in the ideal case. Nevertheless, the potentialities of the implemented pipeline have not been fully exploited in these experiments; in fact, for the considered datasets, the effective usage of the second model in the pipeline is very low (the application of a majority voting or a cosine-based classifier has the same effect).

Regarding the possibility of reducing the number of qubits required to solve a problem by introducing the quantum $k$-NN as a preliminary step, the following relationship holds for the chosen quantum models (it is obtained from Equations~\ref{eq:qknn-qubits-num} and \ref{eq:qbc-qubits-num}):
\begin{equation}
    \label{eq:qknn-qbc-qubits-rel}
    qubits_{qknn} \leq qubits_{qbc} \iff qubits_{features} \leq 3\,.    
\end{equation}
Basically, the application of the implemented quantum $k$-NN as a preliminary step of the quantum binary classifier is advantageous (not disadvantageous) in the number of qubits if and only if the number of data features is less than 5 (less than 9). However, it is worth remarking that \cref{eq:qknn-qbc-qubits-rel} is related to the specific quantum $k$-NN variant and subsequent quantum model that have been employed in this work. Furthermore, the introduction of the quantum $k$-NN has turned out to be advantageous in terms of performance, as shown in \cref{subsubsec:q-pipeline-qbc-comp}.

Another thing worth mentioning regards the unit-norm normalization and, hence, all quantum models exploiting the amplitude encoding of data. In detail, two instances characterized by the same ratio between feature values but different norms (e.g., all features have the same value, but this value is different for the two instances) are normalized to the same data vector. Obviously, this is a significant information loss issue. A possible workaround (not applied in this work) consists in the introduction, before the normalization, of an additional feature related to the norm.

Given the results obtained, the next step will be the development of a version of the quantum $k$-nearest neighbors algorithm more robust to probability fluctuations. Ideally, the quantum $k$-NN in question should produce the nearest neighbors quantum states as output without repeated executions of the same circuit due to the need of estimating some probability value through measurement operations. In this way, it would be possible to realize unified pipelines with other QML models without the need to break the pipeline circuit into distinct segments (the quantum $k$-NN output would be directly provided as input to the subsequent model). Other features of interest are the exploitation of the Euclidean distance as distance metric (it performs better than the cosine distance, as shown in \cref{subsubsec:dist-comp}), the amplitude encoding of the distance values, the absence of oracles, and a low qubit usage. The new quantum $k$-NN variant will also be integrated with more complex QML models, such as the quantum SVM \citep{quantum_svm}, since its classical counterpart has seemed to perform better than the binary classifier as second model in the pipeline. Eventually, more complex datasets will be taken into account, as, most times, extracting the nearest neighbors already identifies the class labels in the datasets used here.


\acks{This work was supported by Q@TN, the joint lab between University of Trento, FBK-Fondazione Bruno Kessler, INFN-National Institute for Nuclear Physics and CNR-National Research Council. In addition, the authors gratefully acknowledge the Italian Ministry of University and Research (MUR), which, under the initiative "Dipartimenti di Eccellenza 2018-2022 (Legge 232/2016)", has provided the computational resources used in the experiments.}



\vskip 0.2in
\bibliography{bibliography}

\end{document}